\documentclass{ptephy_v1}%%%%where ptephy_v1 is the template name
\usepackage{amsmath, amsfonts, amssymb, amsthm}
\usepackage{graphicx}
\usepackage{here}
\usepackage{color}
\begin{document}
\title{Search for sterile neutrinos by shower events
at a future neutrino telescope}
\author{Yabin Wang and Osamu Yasuda}
\affil{Department of Physics, Tokyo Metropolitan University,\\
Hachioji, Tokyo 192-0397, Japan}

\begin{abstract}
It is pointed out that searching for sterile neutrinos at high energy
regions at a future IceCube-like facility has advantages compared with
that of reactor or short baseline accelerator neutrino experiments, in
which the size of the detector and energy resolution make it difficult
to gain good sensitivity for a large mass-squared difference.
In this study we show that it is possible to improve constraints on
sterile neutrino mixing $\theta_{14}$ for 1 eV$^2~\lesssim \Delta
m_{41}^2\lesssim$ 100 eV$^2$ by looking for a dip in the shower events
at an IceCube-like neutrino telescope whose volume is at least 10
times as large as that of IceCube and duration is 10 years.  We also
give an analytic expression for the oscillation probabilities in two
cases where the condition of one mass scale dominance is satisfied.
\end{abstract}
\maketitle

\section{Introduction}
\label{Sec1}

Since Super-Kamiokande first gave firm evidence of neutrino
oscillations in 1998, studying neutrino oscillations due to non-zero
neutrino mass has become a promising direction of exploring new
physics beyond the Standard Model of particle physics.  According to
LEP\,\cite{ParticleDataGroup:2020ssz}, there are only three light
active neutrinos $(\nu_e, \nu_\mu, \nu_\tau)$.  Based on atmospheric and
solar neutrino experiments as well as reactor and accelerator
experiments, there are three mass eigenstates $(\nu_1, \nu_2,
\nu_3)$ with corresponding masses $(m_1, m_2, m_3)$.  The three-flavor eigenstates
of neutrinos are mixed with the three mass eigenstates via a $3 \times
3$ unitary matrix, which is called the PMNS matrix.  In the standard
three-flavor framework of neutrino oscillations, there are six parameters:
three mixing angles, $\theta_{12}$, $\theta_{13}$, and $\theta_{23}$,
two independent mass-squared differences $\Delta m_{21}^2$ and
$\Delta m_{32}^2$, where $\Delta m_{ij}^2 = m_i^2 - m_j^2 \ (i>j; i,
j=1, 2, 3)$, and one CP phase, $\delta$.  As for the problem of the
neutrino mass hierarchy, if $m_3 > m_2 > m_1$ we call it a normal
ordering (NO), whereas, if $m_3 < m_1 < m_2$, it is called an inverted ordering (IO).
Furthermore, based on atmospheric and solar neutrino experiments, there
are two mass-squared differences: $\Delta{m}^2_{\mbox{\rm\scriptsize
    atm}} \equiv |\Delta{m}^2_{32}| \simeq |\Delta{m}^2_{31}|$ and
$\Delta{m}^2_{\mbox{\rm\scriptsize sol}} \equiv \Delta{m}^2_{21}$.
Therefore, we have a hierarchy in the mass-squared differences:
\begin{align}
|\Delta{m}^2_{31}| \simeq 
|\Delta{m}^2_{32}| \gg 
\Delta{m}^2_{21} \,.
\end{align}
The updated values of the above mentioned parameters are as
follows\,\cite{ParticleDataGroup:2020ssz}:
\begin{align}
\sin^2{\theta_{12}} &= 0.307 \pm 0.013 \,, 
\label{osc-param1}\\
\sin^2{\theta_{13}} &= (2.18 \pm 0.07) \times 10^{-2} \,, 
\label{osc-param2}\\
\sin^2{\theta_{23}} &= 0.545 \pm 0.021 \  (\mathrm{NO}) \,, 
\label{osc-param3}\\
\sin^2{\theta_{23}} &= 0.547 \pm 0.021 \ (\mathrm{IO}) \,,  
\label{osc-param4}\\
\Delta{m}^2_{21} &= (7.53 \pm 0.18) \times 10^{-5} \ \mathrm{eV}^2 \,,
\label{osc-param5}\\
\Delta{m}^2_{32} &= (2.453 \pm 0.034) \times 10^{-3} \ \mathrm{eV}^2 \ (\mathrm{NO}) \,, 
\label{osc-param6}\\
\Delta{m}^2_{32} &= (-2.546^{+ 0.034}_{-0.040}) \times 10^{-3} \ \mathrm{eV}^2 \ (\mathrm{IO}) \,,
\label{osc-param7}\\
\delta &= 1.36 \pm 0.17 \ \pi \ \mathrm{rad} \,.
\label{osc-param8}
\end{align}

The first experimental anomaly that cannot be explained by the
standard three-flavor scheme was given by LSND\,\cite{Aguilar:2001ty}.
LSND was looking for an oscillation signal in the channel of
$\bar\nu_\mu \to \bar\nu_e$ and observed an excess of $\bar\nu_e$
events above the background.  This result was supported by the
MiniBooNE experiment\,\cite{MiniBooNE:2018esg}.  This anomaly can be
solved if there is at least one extra neutrino, called a
sterile neutrino (see, e.g., \cite{Abazajian:2012ys}), with a mass-squared difference of $O(1) \ \mathrm{eV}^2$,
which is much larger than $\Delta{m}^2_{\mbox{\rm\scriptsize atm}}$ and
$\Delta{m}^2_{\mbox{\rm\scriptsize sol}}$.

Moreover, solar and reactor neutrino experiments showed that
there are the so-called ``gallium neutrino
anomaly''\,\cite{Giunti:2010zu} and ``reactor antineutrino
anomaly''\,\cite{Mention:2011rk,Huber:2011wv}, in which the observed $\nu_e$
and reactor $\bar\nu_e$ event numbers are smaller than the theoretical
prediction, and these anomalies may be solved by
introducing the existence of light sterile neutrinos whose mass is in
the order of 1 eV, which also motivates us to search for sterile
neutrinos.\,\footnote{The result in Ref.\,\cite{Adey:2019ywk}
  disfavors the new flux of reactor
  antineutrinos\,\cite{Mention:2011rk,Huber:2011wv}, which provided the
  original motivation for reactor antineutrino anomaly from the shape
  analysis of the energy spectrum.  However, as far as the overall
  normalization of reactor antineutrinos is concerned, there is
  uncertainty which is as large as that of the difference between the
  old and new flux\,\cite{Hayes:2013wra} and it is not clear whether
  the reactor antineutrino anomaly is disfavored from the result of
  Ref.\,\cite{Adey:2019ywk}.}

While the combined results on the parameter
space of $(\sin^2{2\theta_{\mu e}}, \Delta{m}^2_{41})$,
where $\sin^2{2\theta_{\mu e}} = 4 |U_{\mu 4}|^2
|U_{e 4}|^2$, $U_{e 4}$ and $U_{\mu 4}$ are the elements of the PMNS
matrix, given by
DayaBay+Bugey-3 ($\bar\nu_e \to \bar\nu_e$) and MINOS ($\nu_\mu \to
\nu_\mu$)\,\cite{MINOS:2020iqj} conflict with that of LSND,
because of the uncertainty in the theoretical calculation of
the reactor antineutrino flux\,\cite{Hayes:2013wra},
the constraint in the
$(\sin^2{2\theta_{\mu e}}, \Delta{m}^2_{41})$-plane in
Ref.\,\cite{MINOS:2020iqj} may not be as strong as they claimed.
This encourages us to improve constraints on sterile neutrino mixing.

Observing high energy neutrinos at IceCube is
advantageous to search for sterile neutrinos.  Because of the matter
effect, there can be a ``dip'' in the neutrino oscillation probability
at certain neutrino energy for the disappearance channel of muon
neutrinos.\,\footnote{See
  Refs.\,\cite{Nicolaidis:1999be,Yasuda:2000xs,Nunokawa:2003ep} for
  earlier references.}
By looking for the dip in the energy spectrum, we can constrain
the upper and lower bounds of the mass-squared difference,
$\Delta{m}^2_{41}$.  Searching for sterile neutrinos at high energy
regions at IceCube has advantages compared with that of lower energy
experiments, such as reactor or short baseline accelerator neutrino
experiments, in which the size of the detectors and energy
resolution make it difficult to gain good sensitivity for a large mass-squared difference.
In this article, {\color{black}assuming the absence of the dip in the energy spectrum,}
we study the sensitivity to
sterile neutrino oscillations using shower events at IceCube rather
than track events studied in Ref.\,\cite{IceCube:2020tka}.

The rest of this article is organized as follows.  In Section \ref{Sec2}, we
briefly introduce the (3+1)-scheme and IceCube experiment.  In
Section \ref{Sec3}, we present some analytical formulae for oscillation
probabilities.  In Section \ref{Sec4}, we perform a numerical analysis of
shower events expected to be observed at a future
IceCube-like facility.  Finally, we draw our conclusions in Section
\ref{Sec5}.  In Appendices \ref{appendix-a} - \ref{appendix-e},
we give some details of deriving analytical expressions
of oscillation probabilities.

\section{Preliminaries}
\label{Sec2}

\subsection{(3+1)-scheme}
\label{Subsec2.1}
The simplest extension of the standard
three-flavor framework assumes that there is only one sterile
neutrino.
The mixing between neutrino flavor eigenstates $(\nu_e, \nu_\mu,
\nu_\tau, \nu_s)$ and mass eigenstates $(\nu_1, \nu_2,
\nu_3, \nu_4)$ is formulated as
\begin{align}
\left( \begin{array}{rl}
\nu_e     \\
\nu_\mu \\
\nu_\tau \\
\nu_s 
\end{array} \right)
=
\left( \begin{array}{cccc}
U_{e1} & U_{e2} & U_{e3} & U_{e4}                  \\
U_{\mu1} & U_{\mu2} & U_{\mu3} & U_{\mu4} \\
U_{\tau1} & U_{\tau2} & U_{\tau3} & U_{\tau4} \\   
U_{s1} & U_{s2} & U_{s3} & U_{s4}    
\end{array} \right)
\left( \begin{array}{rl}
\nu_1 \\ 
\nu_2 \\ 
\nu_3 \\ 
\nu_4   
\end{array} \right) \,,
\end{align}
where $\nu_s$ represents the flavor eigenstate of sterile neutrinos,
$\nu_4$ denotes the fourth mass eigenstate, and $U_{\alpha i}
\ (\alpha=e, \mu, \tau, s; \ i=1, 2, 3, 4)$ denotes the element of the
PMNS matrix $U$.
As the (2+2)-scheme has been excluded by the atmospheric and
solar neutrino data\,\cite{Maltoni:2004ei},
we only consider the (3+1)-scheme in this study.
In this scheme, the mass $m_4$ of the fourth mass eigenstate
is much larger than those of other mass eigenstates.
Therefore, we have
\begin{align}
\Delta{m}^2_{41}    \simeq 
\Delta{m}^2_{42}    \simeq 
\Delta{m}^2_{43}    \gg 
|\Delta{m}^2_{31}| \simeq 
|\Delta{m}^2_{32}| \gg 
\Delta{m}^2_{21} \,.
\end{align}
Further, because the fourth mass eigenstate $\nu_4$ is weekly
mixed with the active three-flavor eigenstates $(\nu_e, \nu_\mu,
\nu_\tau)$, we have the following constraint condition
\begin{align}
|U_{\alpha 4}|^2 \ll 1 
\qquad 
(\alpha = e, \mu, \tau) \,.
\end{align}
In this study, we adopt the following
parametrization of U:
\begin{align}
    U &=
    {\cal R}_{34}(\theta_{34} ,\, \delta_3) \;
    {\cal R}_{24}(\theta_{24} ,\, 0) \;
    {\cal R}_{14}(\theta_{14} ,\, \delta_2) \;
    {\cal R}_{23}(\theta_{23} ,\, 0) \;
    {\cal R}_{13}(\theta_{13} ,\, \delta_1) \; 
    {\cal R}_{12}(\theta_{12} ,\, 0)
\label{u}\\
  &[{\cal R}_{ij}(\theta_{ij},\ \delta_{l})]_{pq} =
  \left\{
  \begin{array}{ll}
    \cos \theta_{ij} & p=q=i,j \\
1 & p=q \not= i,j \\
\sin \theta_{ij} \ e^{-i\delta_{l}} &   p=i;q=j \\
-\sin \theta_{ij} \ e^{i\delta_{l}} & p=j;q=i \\
0 & \mbox{\rm otherwise.}
\end{array}\right.
\label{3+1param2}
\end{align}
Propagation of the neutrino and ani-neutrino flavor eigenstates
$\Psi$ and $\bar\Psi$ in matter can be described by
\begin{align}
i \frac{d\Psi}{dt}
&=
\left(
U{\cal E}
U^{-1}+{\cal A}
\,\right)\,\Psi
\label{DiracEq}\\
i \frac{d\bar\Psi}{dt}
&=
\left(
U^\ast{\cal E}
U^{\ast-1}+{\cal \bar{A}}
\,\right)\,\bar\Psi
\label{DiracEq-anti}\\
\Psi&\equiv
\left( \begin{array}{rl}
\nu_e     \\
\nu_\mu \\
\nu_\tau \\
\nu_s 
\end{array} \right)\,,%\hspace*{-1.7cm}
\quad\bar\Psi\equiv
\left( \begin{array}{rl}
\bar\nu_e     \\
\bar\nu_{\mu} \\
\bar\nu_{\tau} \\
\bar\nu_s 
\end{array} \right)
\\
{\cal E}&\equiv\mbox{\rm diag}
\left(0, \Delta E_{21}, \Delta E_{31}, \Delta E_{41}
\right)
\label{dele}\\
     {\cal A}&\equiv\mbox{\rm diag}
     \left(A_e-iB, -iB, -iB, -A_n
\right)
\nonumber\\
&\equiv\mbox{\rm diag}
     \left(A_e-iB, -iB, -iB, A_s
\right)\\
{\cal \bar{A}}&\equiv\mbox{\rm diag}
\left(-A_e-iB, -iB, -iB, A_n
\right)
\nonumber\\
&\equiv\mbox{\rm diag}
\left(-A_e-iB, -iB, -iB, -A_s
\right)\\
  A_e &= \sqrt{2} \,G_F N_e
  = \left[\frac{\rho}
    {2.6 ~\mathrm{(g \cdot cm^{-3})}}\right]
  \cdot \left(\frac{Y_e}{0.5}\right)
    \cdot 1.0 \times 10^{-13} \ \mathrm{eV}
\\
A_s &\equiv -A_n = \frac{1}{\sqrt{2}} \,G_F N_n
{\color{black}= \left[\frac{\rho}
    {2.6 ~\mathrm{(g \cdot cm^{-3})}}\right]
  \cdot \left(\frac{1-Y_e}{0.5}\right)
    \cdot 5.0 \times 10^{-14} \ \mathrm{eV}}\\
  B &= \frac{1}{2} \,\sigma_{tot} \,(N_p+N_n)\,,
\label{b}
\end{align}
where $\Delta E_{jk} {\color{black}\equiv E_j - E_k
\equiv \sqrt{p^2+m^2_j}-\sqrt{p^2+m^2_k}}\simeq\Delta{m}^2_{jk} / (2E)$,
$E$ {\color{black}($p$)} represents the energy {\color{black}(momentum)} of neutrinos,
$A_e$ represents
the matter effect induced by the charged current (CC) interaction
between $\nu_e$ and electrons, and $A_n\equiv -A_s$ represents the matter
effect induced by the neutral current (NC) interaction of 
$\nu_{\alpha} \ (\alpha = e, \mu, \tau)$ with electrons, $u$ and $d$ quarks.
Note that a $4\times 4$ matrix $(E_1+A_n)\,{\bf 1}_4$
($(E_1-A_n)\,{\bf 1}_4$),
which is proportional to the $4\times 4$ identity matrix ${\bf 1}_4$,
was subtracted from the right hand side of
Eq.\,(\ref{DiracEq}) (Eq.\,(\ref{DiracEq-anti})),
because it affects only the phase of the
probability amplitude $A(\nu_\alpha\to\nu_\beta)$
($A(\bar\nu_\alpha\to\bar\nu_\beta)$) and does
not affect the oscillation probability
$P(\nu_\alpha\to\nu_\beta)$
($P(\bar\nu_\alpha\to\bar\nu_\beta)$).
$G_F = 1.17 \times 10^{-5} \ \mathrm{GeV}^{-2}$ is the
Fermi constant, $N_e$ is the number of electrons in the Earth per
unit volume, which is approximately equal to the number $N_n$ of
neutrons in matter per unit volume, $\rho$ is the
density of matter, and $Y_e=N_p/(N_p+N_n)$ is the
relative number density of electrons in matter,
where $N_p$, which is equal to $N_e$ because an atom
is electrically neutral,
represents the number of protons per unit volume.
$B$ represents the absorption effect\,\cite{Naumov:2001ci} and is
related by the total neutrino--nucleon cross section $\sigma_{tot}$ by 
Eq.\,(\ref{b}), which leads to the expected result
$P(\nu_\alpha\to\nu_\alpha)=\exp[-\sigma_{tot}\,(N_p+N_n)\,L]$
in the absence of neutrino mixing.
{\color{black}The numerical value of the total neutrino--nucleon cross section
is $\sigma_{tot}\sim 10^{-35}$cm$^2$ at $E=1$ TeV, and the contribution
of the CC and NC interactions to $\sigma_{tot}=\sigma_{CC}+\sigma_{NC}$
is shown in Figure \ref{fig3}.}

In vacuum, if we adopt the condition of ``one mass scale
dominance'', which represents the case where only the terms including
$\Delta{m}^2_{41}$ dominate in the oscillation probability, we have
the following formula of oscillation probability:
\begin{align}
P(\nu_{\alpha} \to \nu_{\beta})
\simeq
P(\bar\nu_{\alpha} \to \bar\nu_{\beta})
\simeq 
\left|
\delta_{\alpha \beta} - 
\sin^2{2\theta_{\alpha \beta}} 
\,\sin^2\left( \frac{\Delta{m}^2_{41} L}{4E} \right)
\right| \,,
\end{align}
where
\begin{align}
\sin^2{2\theta_{\alpha \beta}}
=
4|U_{\alpha 4}|^2
\left| \delta_{\alpha \beta} - |U_{\beta 4}|^2 \right|
\qquad
(\alpha, \beta = e, \mu, \tau, s) \,,
\end{align} 
and $L$ represents the baseline length. 

In the discussions on neutrino oscillations at high energy
($E_{\nu} \gtrsim 1$ TeV) with the baseline length of
$O(10^4)$ km, $|\Delta{m}^2_{\mbox{\rm\scriptsize atm}} L / E|$ and $|\Delta{m}^2_{\mbox{\rm\scriptsize sol}} L /
E|$ are negligible, and the condition of the one mass scale
dominance is satisfied.  As we will see below,
the effect of CP violation is negligible in this case.
In the analytical treatment in Section \ref{Sec3},
we discuss oscillation probability
in the limit of one mass scale dominance.

\subsection{IceCube experiment}
\label{Subsec2.2}

IceCube is a neutrino observatory located at the South Pole.  IceCube
observes atmospheric and astrophysical neutrinos,
whose energy ranges approximately from 100 GeV to above 10$^9$
GeV.  Among the neutrino events, there are track and
shower events.  Track events are recorded by the Cherenkov light
of muons produced by the CC interaction
between muon neutrinos and nuclei in ice.  Shower events include
electromagnetic and hadronic shower events,
which are produced by the CC interactions of $\nu_e$ and $\nu_\tau$ as
well as the NC interactions of any flavor of
neutrinos.  Track events have good angular resolution but poor
energy resolution, whereas shower events have poor angular
resolution but good energy resolution.

When up-going neutrinos arrive at IceCube after passing through the
Earth, the effective mixing angle in matter can be enhanced due to the
matter effect and a dip can appear in the neutrino oscillation
probability for the disappearance channel.  By searching for the dip
in the oscillation probability, we can give the lower and
upper bounds on the third mass-squared difference $\Delta{m}^2_{41}$.
This was done for the disappearance channel of muon
neutrinos\,\cite{IceCube:2020tka}.

At high energy ($\gtrsim 10$ GeV) muon neutrinos constitute most
atmospheric neutrinos because most high energy atmospheric
muons do not decay into electrons and neutrinos before they reach the
Earth.  Therefore, at high energy the flux of atmospheric
$\nu_e+\bar\nu_e$, which mainly come from kaon decays, is much
smaller than that of $\nu_\mu+\bar\nu_\mu$.  Despite this fact,
motivated by the reactor antineutrino and gallium neutrino
anomalies, we study the sensitivity of shower events to neutrino
oscillations, anticipating the large detector volume may improve this
disadvantage in the future.

In the energy range of 1-100 TeV, prompt neutrinos, which come
from the decay of charmed mesons, may have fluxes of $\nu_\mu$ and
$\nu_e$, comparable to that of the conventional atmospheric
neutrinos.  Although prompt neutrinos can be essential components
of atmospheric neutrinos, the result of Ref.\,\cite{IceCube:2015mgt}
agrees with a hypothesis of zero prompt neutrino event.
Therefore, in the analysis in Section \ref{Sec4}, we do
not consider the contribution of prompt neutrinos in the search for
sterile neutrinos.

\section{Some analytical treatment}
\label{Sec3}

In the numerical analysis in Section \ref{Sec4}, it is useful to have
analytic expressions for oscillation probabilities to understand the
qualitative behavior of the energy spectrum etc.  So in this section
we discuss some analytic expressions for oscillation probabilities
in matter with constant density.
For simplicity we discuss a case without absorption, i.e.,
$B=0$ in Eqs.\,(\ref{DiracEq}) and (\ref{DiracEq-anti}), since the
analytic expressions for oscillation probabilities with $B\ne 0$
are complicated.

\subsection{Atmospheric neutrinos}
\label{Subsec3.1}

In this subsection, we discuss the oscillation probabilities of
atmospheric neutrinos.  Upward-going atmospheric
neutrinos pass through the Earth and are observed at the IceCube
detector.  The baseline length $L$ is given by
\begin{align}
  \hspace*{-20mm}
  L = -R\cos\Theta +\sqrt{(R+h)^2-(R\sin\Theta)^2}\,,
\end{align}
where $R$ is the Earth's radius,
$\Theta$ is the zenith angle, and $h\sim$ 20 km
is the thickness of the atmosphere.

In this study, we discuss neutrinos at high energy ($E_{\nu} \gtrsim 1$
TeV) with a baseline length $L\lesssim 13000)$ km.  Since
$|\Delta{m}^2_{\mbox{\rm\scriptsize atm}} L / E|\ll 1$ and
$|\Delta{m}^2_{\mbox{\rm\scriptsize sol}} L /E|\ll 1$, the condition of
the one mass scale dominance is satisfied.  In this case,
the energy difference matrix in Eq.\,(\ref{dele}) can
be approximated as
\begin{eqnarray}
\hspace*{-45mm}{\cal E}\simeq \mbox{\rm diag}
\left(0, 0, 0, \Delta E
\right)
\,,
\end{eqnarray}
where we have defined
\begin{eqnarray}
&{\ }&\hspace*{-80mm}
\Delta E\equiv\Delta E_{41}
\end{eqnarray}
for simplicity, and
the matrices
${\cal R}_{23}(\theta_{23} ,\, 0)$,
${\cal R}_{13}(\theta_{13} ,\, \delta_1)$,
${\cal R}_{12}(\theta_{12} ,\, 0)$
disappear from $U$ in Eqs.\,(\ref{DiracEq}) and (\ref{DiracEq-anti}).

It turns out that the two cases
$\theta_{14} \ne 0$, $\theta_{24} = \theta_{34} = 0$
and $\theta_{14} = 0$, $\theta_{24} \ne 0$, $\theta_{34} \ne 0$
in the limit of one mass scale dominance
can be reduced to a two-flavor problem.
The former (latter) has a resonance
in the neutrino (antineutrino) mode,
so we discuss these two cases separately.
In both cases, CP phases do not appear in the
oscillation probability, as in
the two-flavor case.

\subsubsection{The case with $\theta_{14} \ne 0$,
 and $\theta_{24} = \theta_{34} = 0$}
\label{Subsubsec3.1.1}
We first consider a case where there is only one non-zero mixing
angle $\theta_{14}$, whereas $\theta_{24}$ and $\theta_{34}$ vanish.  In
this case, there is mixing only between electron and
sterile neutrinos, and Eq.\,(\ref{DiracEq}) reduces to the
one in the two-flavor case.  From Eq.\,(\ref{H1.2}) in
Appendix\,\ref{appendix-a}, we can calculate the oscillation
probability of $\nu_e \to \nu_e$:
\begin{align}
  \hspace*{-20mm}
P(\nu_e \to \nu_e)
=
1 - \sin^22\tilde\theta_{14}\,
\sin^2\left(\frac{\Delta\tilde{E}_1 L}{2}\right) \,,
\label{P1}
\end{align}
where $\tilde\theta_{14}$ and $\Delta\tilde{E}_1$ are, respectively, defined by
\begin{align}
  \hspace*{-35mm}
\tan{2\tilde\theta_{14}} 
&= 
\frac{\Delta E \,\sin{2\theta_{14}}}
     {\Delta E \cos{2\theta_{14}} - A_e + A_s}  \,.
     \label{theta14tilde}\\
     \Delta\tilde E_1
&= 
     \sqrt{(\Delta E \cos{2\theta_{14}} - A_e + A_s)^2
       + (\Delta E \sin{2\theta_{14}})^2} \,.
\label{etilde1}
\end{align}
Since $A_e - A_s\simeq A_e/2 > 0$
and $\Delta E = \Delta{m}^2_{41}/(2E) > 0$,
Eq.\,(\ref{theta14tilde}) indicates
that a resonance occurs only in the
neutrino mode at the neutrino energy
$E = {\color{black}\Delta{m}^2_{41}\cos{2\theta_{14}}/\{2(A_e - A_s)\}
\simeq}$ $\Delta{m}^2_{41}/\{2(A_e - A_s)\}$,
where we have assumed $|\theta_{14}|\ll 1$.

Moreover, it is useful to study the behavior of oscillation
probabilities when considering the following:
\begin{eqnarray}
 &{\ }&\hspace*{-80mm}
\Delta{m}^2_{41} \to +\infty \,.
\label{E}
\end{eqnarray}
If $\Delta{m}^2_{41} \to +\infty$, then
the vacuum oscillation contribution
$\Delta E$ dominates over the matter term $A_e - A_s$
in Eq.\,(\ref{theta14tilde}),
so we have $\tilde\theta_{14} \to \theta_{14}$,
whereas $\Delta\tilde{E}\to +\infty$ leads to rapid oscillation
in Eq.\,(\ref{P1}).  Thus we obtain
\begin{align}
  \hspace*{-40mm}
P(\nu_e \to \nu_e)
\to 1 - \frac{1}{2}\,\sin^22\theta_{14} \,.
\label{P2}
\end{align}
From Eq.\,(\ref{P2}), we observe that if $\Delta{m}^2_{41} \to
+\infty$, the oscillation probability $P(\nu_e \to \nu_e)$
becomes independent of the zenith angle $\Theta$ and neutrino
energy $E$.  The fact that $P(\nu_e \to \nu_e)$ becomes a constant
suppression as $\Delta{m}^2_{41} \to +\infty$ implies that
the significance of the (3+1)-scheme with $\theta_{14}\ne 0$ at very
large $\Delta{m}^2_{41}$ depends entirely on the systematic errors.

\subsubsection{The case with $\theta_{14} = 0$, $\theta_{24} \ne 0$
and $\theta_{34} \ne 0$}
\label{Subsubsec3.1.2}

Next, we consider a case where there are two non-zero
mixing angles $\theta_{24}$ and $\theta_{34}$, whereas $\theta_{14}$
vanishes.
Although the (3+1)-scheme for which we perform
the numerical analysis is not this case,
it is instructive to see how a resonance occurs
in the antineutrino mode in this case.

As explained in Appendix \ref{appendix-b},
there can be a resonance in the
antineutrino mode (see Eq.\,(\ref{13.3}).).
Therefore, we discuss the
disappearance channel $\bar\nu_\mu \to \bar\nu_\mu$,
introducing a new set of angles
\begin{eqnarray}
  &{\ }&\hspace*{-70mm}
\tan{\varphi_{12}} 
\equiv 
\displaystyle\frac{t_{34}}{s_{24}}\,, 
\label{phi12}\\
  &{\ }&\hspace*{-70mm}
\sin{\varphi_{13}}
\equiv c_{24}c_{34}\,, 
\label{phi13}\\
  &{\ }&\hspace*{-70mm}
\tan{\varphi_{23}} 
\equiv 
\displaystyle\frac{s_{34}}{t_{24}} \,,
\label{13.1}
\end{eqnarray}
where $c_{jk}\equiv\cos\theta_{jk}$,
$s_{jk}\equiv\sin\theta_{jk}$,
$t_{jk}\equiv\tan\theta_{jk}$, and
$\theta_{jk}~(j,k=1,\cdots,4)$ is the original
mixing angle in the parametrization (\ref{u}).
The oscillation probabilities $P(\bar\nu_\mu \to \bar\nu_\mu)$
and $P(\bar\nu_\mu \to \bar\nu_\tau)$ can be expressed as:
\begin{align}
P(\bar\nu_\mu \to \bar\nu_\mu)
= {} &
1 -
C_{23}^4\sin^22\tilde\varphi_{13}\,
\sin^2\left(\frac{\Delta\tilde{E}_2 L}{2}\right)
\nonumber\\
&~~-\sin^22\varphi_{23}\,
\left\{\tilde{S}_{13}^2
\sin^2\left(\frac{\varepsilon_{2 -} L}{2}\right)
+\tilde{C}_{13}^2 
\sin^2\left(\frac{\varepsilon_{2 +} L}{2}\right)
\right\}\,,
\label{P3}\\
P(\bar\nu_\mu \to \bar\nu_\tau)
={} &
\sin^22\varphi_{23}
  \left\{
  \tilde{S}_{13}^2
\sin^2\left(\frac{\varepsilon_{2 -} L}{2}\right)
+\tilde{C}_{13}^2 
\sin^2\left(\frac{\varepsilon_{2 +} L}{2}\right)
\right.\nonumber\\
&\hspace*{20mm}-\left.\frac{1}{4}
\sin^22\tilde\varphi_{13}\,
\sin^2\left(\frac{\Delta\tilde{E}_2 L}{2}\right)
\right\}\,,
\label{P4}
\end{align}
where $\tilde\varphi_{13}$,
$\Delta\tilde{E}_2$ and $\varepsilon_{2 \mp}$
are, respectively, given by
\begin{align} 
\tan{2\tilde\varphi_{13}}
&=
\frac{\Delta E \sin{2\varphi_{13}}}
{\Delta E \cos{2\varphi_{13}} + A_s}\,,
\label{13.2}\\
\Delta\tilde{E}_2 
&=
\sqrt{(\Delta E \cos{2\varphi_{13}} + A_s)^2
  + (\Delta E \sin{2\varphi_{13}})^2} \,,
\label{epsilon2.1}
\\
\varepsilon_{2 \mp}
&=
\frac{\Delta E - A_s}{2}
\mp
\frac{1}{2} \Delta\tilde{E}_2 \,,
\label{E2.1}
\end{align}
and we introduced the notations
$\tilde{S}_{13} = \sin{\tilde\varphi_{13}}$, 
$\tilde{C}_{13} = \cos{\tilde\varphi_{13}}$,
$S_{23} = \sin{\varphi_{23}}$
and
$C_{23} = \cos{\varphi_{23}}$.
If $\theta_{24}$ and $\theta_{34}$ are
both small so that
$0<s_{24}\ll 1$ and $0<s_{34}\ll 1$ are satisfied,
from Eq.\,(\ref{phi13}), we have
$0<\pi/2 - \varphi_{13}\ll 1$,
which implies that $2\varphi_{13}\simeq \pi$
and $\cos2\varphi_{13}\simeq -1$, so that
the resonance condition, in which 
the denominator of Eq.\,(\ref{13.2})
vanishes, is possible only for antineutrinos.
Notice that the denominator for neutrinos
would be $\Delta E \cos2\varphi_{13} - A_s$,
which is negative for any
value of the neutrino energy $E>0$.

Now, let us consider the behavior of the oscillations probabilities
$\bar\nu_\mu \to \bar\nu_\mu$ and $\bar\nu_\mu \to \bar\nu_\tau$
as $\Delta{m}^2_{41} \to +\infty$, where
$\Delta E$ dominates over the matter term $A_s$
in Eq.\,(\ref{13.2}),
so we have $\tilde\varphi_{13} \to \varphi_{13}$.
Moreover, for the three energy differences,
two of them become large
whereas one remains finite:
\begin{align}
\Delta\tilde{E}_2 
& \sim
\Delta E\to
+\infty \,.
\label{E2.2}\\
\varepsilon_{2 +}
&=
\frac{\Delta E - A_s}{2}
+
\frac{1}{2} \Delta\tilde{E}_2
\sim
\Delta E %+ A_s \sin^2{\varphi_{13}} 
\to
+\infty \,.
\label{epsilon2.3}\\
\varepsilon_{2 -}
&=
\frac{\varepsilon_{2 +}\varepsilon_{2 -}}{\varepsilon_{2 +}}
\nonumber\\
&=\frac{A_s\Delta E\cos^2\varphi_{13}}{\varepsilon_{2 +}}
\to
A_s \cos^2{\varphi_{13}} \,,
\label{epsilon2.2}
\end{align}
From this, $\sin^2(\Delta\tilde{E}_2L/2)$
and $\sin^2(\varepsilon_{2 +}L/2)$ are averaged
to yield $1/2$, whereas $\sin^2(\varepsilon_{2 +}L/2)$
becomes $\sin^2(A_s \cos^2{\varphi_{13}}L/2)$,
which does not depend on the neutrino energy
$E$ but depends on the baseline length $L$.
Thus, we obtain 
\begin{align}
P(\bar\nu_\mu \to \bar\nu_\mu)
&\to 
1 -
\frac{C_{23}^4}{2}\,\sin^22\varphi_{13}
\nonumber\\
&~~-\sin^22\varphi_{23}\,
\left\{{S}_{13}^2
\sin^2\left(\frac{A_s \cos^2{\varphi_{13}} L}{2}\right)
+\frac{{C}_{13}^2 }{2}
\right\}\,,
\label{P5}\\
P(\bar\nu_\mu \to \bar\nu_\tau)
&\to
\sin^22\varphi_{23}
  \left\{
  {S}_{13}^2
\sin^2\left(\frac{A_s \cos^2{\varphi_{13}} L}{2}\right)
+\frac{{C}_{13}^2 }{2}
-
\frac{1}{8}
\sin^22\varphi_{13}
\right\}\,,
\label{P6}
\end{align}
Eq.\,(\ref{P5}) shows that,
even as $\Delta{m}^2_{41} \to +\infty$,
the disappearance probability
has some dependence on the zenith angle $\Theta$
(not only $P(\bar\nu_\mu \to \bar\nu_\mu)$
but also $P(\nu_\mu \to \nu_\mu)$ because
Eq.\,(\ref{P5}) is an even function of $A_s$.),
which is why the IceCube has some
sensitivity to $\theta_{24}$ even at
$\Delta m^2_{41}=100$ eV$^2$
(cf. the upper panel of FIG.19
in Ref.\,\cite{IceCube:2020tka}).
This phenomenon has also been known
in the standard three-flavor
scenario\,\cite{Fogli:1997ff,Foot:1998pv},
where the atmospheric neutrino
data was accounted for to some extent,
even for a low value of the mass-squared difference
$|\Delta{m}^2_{\mbox{\rm\scriptsize atm}}|$.

\subsection{Astrophysical neutrinos}
\label{Subsec3.2}

Next, we discuss neutrino oscillations of astrophysical neutrinos.  We
consider the following propagation process: astrophysical neutrinos
start from the source and travel a long distance in outer space (in
vacuum), then reach and pass through the Earth (in matter), and are finally
observed at IceCube.  In the following, we develop formulae for
calculating the oscillation probabilities in this process.

Calculating these oscillation probabilities is similar to that of
the day-night effect of solar neutrinos.  We denote $L_0$ as the
astronomical travel distance of astrophysical neutrinos in outer
space, and $L$ as the propagation distance of astrophysical neutrinos
inside the Earth.  Because $|\Delta m^2_{ij}L_0/E|\gg
1~(i,j=1,\cdots,4)$, to a good approximation, we can assume $L_{0} \to
+\infty$.  Further, we denote $T$ as the transition matrix of the
neutrino flavor state inside the Earth.  Then we have the formulae of
oscillation probabilities in the entire propagation process:
\begin{align}
P(\nu_{\alpha} \to \nu_{\beta})
& =
\lim_{L_{0}\to+\infty}\sum_{\rho j}
T_{\beta \rho} \,U_{\rho\, j} \,\exp(-i L_{0} E_j)\, U^{-1}_{j \alpha}
\notag \\
& \qquad\qquad\times
\left( \sum_{\sigma n}
T_{\beta \sigma} \,U_{\sigma n}\, \exp(-i L_{0} E_n)\, U^{-1}_{n \alpha} 
\right)^\ast \notag \\
& =
\lim_{L_{0}\to+\infty}
\sum_{\rho \,\sigma j n}
T_{\beta \rho}\, T_{\beta \sigma}^\ast
U_{\rho j}\, U_{\sigma n}^\ast
\exp(i L_{0} \Delta{E_{n j}}) 
U_{\alpha j}^\ast U_{\alpha n} \notag \\
& =
\sum_{\rho \,\sigma n}
T_{\beta \rho} \,T_{\beta \sigma}^\ast
U_{\rho n} \,U_{\sigma n}^\ast\,
|U_{\alpha n}|^2 \qquad
(\alpha, \beta = e, \mu, \tau, s) \,,
\label{P7}
\end{align}
where we have used the unitary condition of the PMNS matrix $U$.

\subsubsection{The case with $\theta_{14} \ne 0$, and $\theta_{24} = \theta_{34} = 0$}
\label{Subsubsec3.2.1}
The flavor ratio of neutrinos in the case of
$\theta_{14} \ne 0$, $\theta_{24} = \theta_{34} = 0$
is given by the following
(for a detailed description, see Appendix \ref{appendix-c}):
\begin{align}
F(\nu_e) 
:
F(\nu_\mu) 
:
F(\nu_\tau)
:
F(\nu_s) 
&\simeq 1 - |T_{14}|^2 : 1 : 1: |T_{14}|^2\,,
\label{ratio1}
\end{align}
where $|T_{14}|^2$ is defined as
\begin{align}
|T_{14}|^2 = \sin^22\tilde\theta_{14}
  \sin^2\left(\frac{\Delta\tilde{E}_1L}{2}\right)\,,
\label{t142}
\end{align}
$\tilde\theta_{14}$ and $\Delta\tilde{E}_1$ are defined by
Eqs.\,(\ref{theta14tilde}) and (\ref{etilde1}), respectively, and $L$
is the path length of astrophysical neutrinos inside the Earth.
Note that the factor $|T_{14}|^2$ and hence the
flavor ratio (\ref{ratio1}) have a nontrivial dependence on the
neutrino energy $E$ especially near the resonance region $E
=\Delta m^2_{41}\cos2\theta_{14}/\{2(A_e-A_s)\}
\sim \Delta m^2_{41}/\{2(A_e-A_s)\}$.

\subsubsection{The case with $\theta_{14} = 0$, $\theta_{24} \ne 0$
  and $\theta_{34} \ne 0$}
\label{Subsubsec3.2.2}

In the case with $\theta_{14} = 0$, $\theta_{24} \ne 0$,
and $\theta_{34} \ne 0$,
as described in detail in Appendix \ref{appendix-d},
the flavor ratio of {\it antineutrinos} is expressed as follows:
\begin{align}
F(\bar\nu_e) 
:
F(\bar\nu_\mu) 
:
F(\bar\nu_\tau)
:
F(\bar\nu_s) 
&\simeq
1
:
1 - C_{23}^2 \,|T_{44}|^2
:
1 - S_{23}^2 \,|T_{44}|^2
:
1 - |T_{44}|^2
\,,
\label{ratio2}
\end{align}
where $|T_{44}|^2$ is defined as
\begin{align}
|T_{44}|^2
&= 1 - \sin^22\tilde\varphi_{13}\,
\sin^2\left(\frac{\Delta\tilde{E}_2 L}{2}\right)\,,
\label{t442}
\end{align}
$C_{23}\equiv \cos\varphi_{23}$,
$S_{23}\equiv \sin\varphi_{23}$,
the new mixing angles $\varphi_{ij}~(i,j=1,2,3)$
are defined in Eqs.\,(\ref{phi12}), (\ref{phi13}), (\ref{13.1}),
$\tilde\varphi_{13}$ and $\Delta\tilde{E}_2$ are defined by
Eqs.\,(\ref{13.2}) and (\ref{epsilon2.1}), respectively, and $L$
is the path length of astrophysical neutrinos inside the Earth.
Notice that the factor $|T_{44}|^2$ and therefore the
flavor ratio (\ref{ratio2}) also have a nontrivial dependence on the
neutrino energy $E$ especially near the resonance region
$E{\color{black}=-\Delta m^2_{41}\cos2\varphi_{13}/(2A_s)}
\sim \Delta m^2_{41}/(2A_s)$, where we have assumed $|\pi/2-\varphi_{13}|\ll 1$.

\section{Numerical analysis of shower events}
\label{Sec4}

In this section, for simplicity, we perform numerical analysis to
obtain the sensitivity to $\theta_{14}$ as a function of
$\Delta{m}^2_{41}$ in the (3+1)-scheme, assuming
$\theta_{24}=\theta_{34}=0$.  As shown in Section \ref{Sec3}, the
case with $\theta_{14}\ne 0, \theta_{24}=\theta_{34}=0$ has a
resonance in the neutrino mode, whereas the case with $\theta_{14}= 0,
\theta_{24}\ne 0, \theta_{34}\ne 0$ has one in the antineutrino mode.
If we turn on small $\theta_{24}$ and $\theta_{34}$, the
resonance in the neutrino mode has small perturbation with respect to
$\theta_{24}$ and $\theta_{34}$, and we expect that our conclusion does
not change significantly.

We assume that the numbers of events at an IceCube-like
neutrino telescope are given by extrapolating those in
FIG.6 in Ref.\,\cite{IceCube:2015mgt}, i.e.,
they are given by a factor
(volume of the IceCube-like detector) $\times$
(number of days of measurement) /
[(volume of the IceCube) $\times$ (332 days)].
The figure
shows the energy spectrum of
various numbers of shower events for
332 days at IceCube.  They comprise
conventional atmospheric $\nu_e+\bar{\nu}_e$,
conventional atmospheric $\nu_\mu+\bar{\nu}_\mu$,
(which look like shower events even after
the experimental cuts), 
astrophysical $\nu_\beta+\bar{\nu}_\beta~(\beta=e, \mu, \tau)$,
and cosmic ray muons.
Note that the IceCube collaboration
concluded that zero prompt neutrino event
is consistent with their data, so we
also assume that there is no prompt neutrino event
in our analysis.

\subsection{Numbers of events}
\label{Subsec4.1}

The neutrinos observed at an IceCube-like
detector are either atmospheric or
astrophysical, and are
expected to be produced mainly from
pion and kaon decays.  So, it is expected that
there are only $\nu_e+\bar{\nu}_e$ and
$\nu_\mu+\bar{\nu}_\mu$ at the production point.
However, during neutrino propagation,
the processes
$\nu_\alpha\to\nu_\tau~(\alpha=e, \mu)$
and $\bar{\nu}_\alpha\to\bar{\nu}_\tau~(\alpha=e, \mu)$
exist in general neutrino mixing.
For atmospheric neutrinos,
in the standard three-flavor mixing scenario
or (3+1)-scheme where the only
non-zero sterile neutrino mixing angle is
$\theta_{14}$,
there are very few $\nu_\tau+\bar{\nu}_\tau$
from the conventional atmospheric neutrinos
through the oscillations
$\nu_\mu\to \nu_\tau$ and
$\bar{\nu}_\mu\to \bar{\nu}_\tau$
because the oscillation probability
is very small:
$P(\nu_\mu\to \nu_\tau)
\simeq P(\bar{\nu}_\mu\to \bar{\nu}_\tau)
\simeq \sin^22\theta_{23}\sin^2(\Delta m^2_{31}L/4E) \ll 1$
for $E> 1$ TeV,
$L\lesssim 2R$, where $R$
is the Earth's radius.
Meanwhile, for
astrophysical neutrinos,
we have
$\nu_\tau+\bar{\nu}_\tau$ because of
oscillations with an astronomical baseline
length\,\cite{Learned:1994wg}.

Thus, the number of events
($N_{jk}(\nu_\beta)$ for $\nu_\beta$
and $N_{jk}(\bar{\nu}_\beta)$ for $\bar{\nu}_\beta$)
for the energy bin
($E_j<E<E_{j+1}$) and zenith angle bin
($\cos\Theta_k<\cos\Theta<\cos\Theta_{k+1}$)
can be expressed in terms of the flux
$\Phi(\nu_\alpha, E', \Theta')$,
the cross section\,\cite{Gandhi:1998ri}
($\sigma_X(E')$ for neutrinos
and $\bar{\sigma}_X(E')$ for antineutrinos,
for CC ($X=CC$) and
NC ($X=NC$) interactions,
respectively) and the detection efficiency
$\epsilon_X(E,\nu_\beta)~(X=CC, NC)$ as follows:
\begin{align}
  &{\ }%\hspace*{-10mm}
\left\{
\begin{array}{c}
  N_{jk}(\nu_\beta)\\
N_{jk}(\bar{\nu}_\beta)
\end{array}\right\}
\nonumber\\
  %{\ }&\hspace*{-10mm}
  =&\int_{\log_{10} E_j}^{\log_{10} E_{j+1}}d\log_{10} E\,
\int_{\cos\Theta_j}^{\cos\Theta_{j+1}}d\cos\Theta\,
\int_{-\infty}^{\infty}d\log_{10} E'\,
\int_{-1}^1d\cos\Theta'\,
\nonumber\\
  %{\ }&\hspace*{-5mm}
&\times \,R_e(E, E',\Delta_e(E))\,
R_z(\Theta,\Theta',\Delta_z(E))
\nonumber\\
  %{\ }&\hspace*{-5mm}
&\times\sum_\alpha \sum_{X=CC,NC}\,
\left\{
\begin{array}{c}
\Phi(\nu_\alpha, E', \Theta')\,
P(\nu_\alpha\to\nu_\beta)\,
\sigma_X(E')\,\epsilon_X(E',\nu_\beta)\\
\Phi(\bar{\nu}_\alpha, E', \Theta')\,
P(\bar{\nu}_\alpha\to\bar{\nu}_\beta)\,
\bar{\sigma}_X(E')\,\epsilon_X(E',\bar{\nu}_\beta)
\end{array}\right\}
\label{no-events}
\end{align}
$R_e$ and $R_z$ represent the
resolution functions in the neutrino energy
and zenith angle, respectively:
\begin{align}
  R_e(E, E',\Delta_e(E))&\equiv
  \frac{1}{\sqrt{\pi}\Delta_e(E)}\,
  \exp\left\{-\frac{(\log_{10} E-\log_{10} E')^2}{\Delta_e(E)^2}
    \right\}
\nonumber\\
R_z(\Theta,\Theta',\Delta_z(E))&\equiv
N(\Theta)\,
  \exp\left\{-\frac{(\cos\Theta-\cos\Theta')^2}{\Delta_z(E)^2}
    \right\}
\nonumber\\
N(\Theta)&\equiv\left[\int_{-1}^1d\cos\Theta'\,
  \exp\left\{-\frac{(\cos\Theta-\cos\Theta')^2}{\Delta_z(E)^2}
    \right\}\right]^{-1}
\end{align}
where $E'$ and $\Theta'$
are the true neutrino energy and zenith angle, respectively,
whereas $E$ and $\Theta$ are the reconstructed ones, respectively.
We use the energy resolution $\Delta_e(E)$ and
the zenith angle resolution $\Delta_z(E)$
which are given by the blue curve ``iterative CREDO''
in FIG.5 in Ref.\,\cite{IceCube:2015mgt}.

\subsubsection{Atmospheric neutrinos}

We use the atmospheric neutrino flux
in Ref.\,\cite{Honda:2006qj} in our analysis.
However, the flux in Ref.\,\cite{Honda:2006qj}
is given only up to 10 TeV, and we need
the flux up to ${\cal O}$(100 TeV) for our analysis.
Therefore, we adopt the following empirical formula
in Ref.\,\cite{Chirkin:2004ic} for the
atmospheric neutrino flux, and extrapolate
it up to ${\cal O}$(100 TeV):
\begin{align}
\left\{
\begin{array}{c}
  \Phi^{\mbox{\scriptsize atm}}(\nu_\mu, E, \Theta)\\
  \Phi^{\mbox{\scriptsize atm}}(\bar{\nu}_\mu, E, \Theta)
\end{array}\right\}  
=&\frac{2.85\cdot 10^{-2}}{\mbox{ cm}^2\mbox{ sr}\mbox{ s}\mbox{ GeV}}
\cdot \mbox{A}
\cdot \left(\frac{E}{\mbox{GeV}}\right)^{-\gamma}
      \nonumber\\
&\times\left(\frac{1}
      {1+6 \,E\cos\Theta^\ast /\mathrm{121\ GeV}}
      +\frac{0.213}{1+1.44\, E\cos\Theta^\ast 
        /\mathrm{897\ GeV}}\right)
      \label{phimu}\\
\left\{
\begin{array}{c}
\Phi^{\mbox{\scriptsize atm}}(\nu_e, E, \Theta)\\
\Phi^{\mbox{\scriptsize atm}}(\bar{\nu}_e, E, \Theta)
\end{array}\right\}  
      =&\frac{2.4\cdot 10^{-3}}{\mbox{cm}^2\mbox{ sr}\mbox{ s}\mbox{ GeV}}
      \cdot \mbox{A} \cdot
      \left(\frac{E}{\mbox{GeV}}\right)^{-\gamma}
      \nonumber\\
      &\times \left(\frac{0.05}{1+1.5\, E\cos\Theta^\ast 
        /\mathrm{897\ GeV}}
      +\frac{0.185}{1+1.5\, E\cos\Theta^\ast 
        /\mathrm{194\ GeV}}\right.
            \nonumber\\
      &\left.+\frac{11.4\,E^{\,\zeta(\Theta)}}
      {1+1.21 \,E\cos\Theta^\ast /\mathrm{121\ GeV}}\right)
      \label{phie}
\end{align}
where the cosine of the zenith angle $\Theta$ is modified as
that of the effective one $\Theta^\ast$ by considering
the curvature of the atmosphere:
\begin{align}
  \cos\Theta^\ast&=\sqrt{\frac{\cos^2\Theta+p_1^2
      +p_2\cos^{\,p_3}\Theta+p_4\cos^{\,p_5}\Theta}{1+p_1^2+p_2+p_4}}
  \nonumber\\
  (p_1, p_2, p_3, p_4, p_5)&=
  (0.102573, -0.068287, 0.958633, 0.0407253, 0.817285).
      \nonumber
\end{align}
$A$ is the normalization, given by
$A=0.646$ for $\nu_\mu$, $\bar{\nu}_\mu$ and
$A=0.828$ for $\nu_e$, $\bar{\nu}_e$.
$\gamma$ is the spectral index for atmospheric neutrinos,
and the flux for each
flavor ($\nu_\mu$, $\bar{\nu}_\mu$, $\nu_e$ and $\bar{\nu}_e$)
in Ref.\,\cite{Honda:2006qj}
is best reproduced by
$\gamma = 2.680$ (for $\nu_\mu$),
2.696 (for $\bar{\nu}_\mu$),
2.719 (for $\nu_e$), and
2.741 (for $\bar{\nu}_e$).
Hence, we use Eqs.\,(\ref{phimu}) and (\ref{phie})
with these values for $\gamma$ up to the
neutrino energy $E\lesssim$ 1 PeV.
$\zeta$ in Eq.\,(\ref{phie}) is given by
\begin{align}
\zeta&=a+b\log_{10}(E/\mathrm{GeV})
      \nonumber\\
a&=
\max(0.11-2.4\cos\Theta,
-0.46-0.54\cos\Theta)
      \nonumber\\
b&=
\min(-0.22+0.69\cos\Theta,% \quad & \cos(\Theta)<0.3\\
-0.01+0.01\cos\Theta)% \quad & \cos(\Theta)\geqslant 0.3
      \nonumber
\end{align}

\subsubsection{Detection efficiency for shower events}

The detection efficiency
$\epsilon_X(E_{\color{black}\nu},\nu_\beta)~(\beta=e, \mu; X=CC, NC)$
in Eq.\,(\ref{no-events}) for shower events
can be deduced from the number of shower events
of atmospheric neutrinos given in Ref.\,\cite{IceCube:2015mgt},
if we make the following assumptions:
\begin{enumerate}
\renewcommand{\labelenumi}{(\roman{enumi})}
\item The detection efficiencies for neutrinos
and antineutrinos are the same.
\begin{align}
  &{\ }&\hspace*{-10mm}
  \epsilon_X(E_{\color{black}\nu},\nu_\beta)
  =\epsilon_X(E_{\color{black}\nu},\bar{\nu}_\beta)
  \quad(\beta=e, \mu, \tau;X=CC, NC)
\end{align}

\item The energy dependencies of 
  $\epsilon_{NC}(E_\nu^{NC},\nu_e)$ and $\epsilon_{CC}(E_\nu^{CC},\nu_e)$
  are related by the following formula:
\begin{align}
  &{\ }&\hspace*{-50mm}
\color{black}\epsilon_{NC}(E_\nu^{NC},\nu_e) = \mathrm{const.}\,\epsilon_{CC}(E_\nu^{CC},\nu_e)
\end{align}
{\color{black}where the neutrino energy $E_\nu^{CC}$ ($E_\nu^{NC}$) in the charged
(neutral) current interaction are related through the observed energy
$E_{obs}$ by the following:
\footnote{\color{black}
  Since the contribution of the neutral current interaction to
  the numbers of events is not dominant, the precise value
  $E_\nu-E_{obs}$ in the relation
  between the neutrino energy $E_\nu^{CC}$ ($E_\nu^{NC}$) and
  the observed energy $E_{obs}$ does not affect the sensitivity
  to $\theta_{14}$ very much.}}
\begin{eqnarray}
  &{\ }&\hspace*{-15mm}
  \color{black}E_{obs}\simeq E_{\nu}^{CC}-1~\mbox{\rm GeV}
%  \nonumber\\
%  &{\ }&\hspace*{-10mm}
       \simeq E_{\nu}^{NC}-10~\mbox{\rm GeV}\,.
\label{enucc}
\end{eqnarray}

\item The detection efficiencies $\epsilon_{NC}(E,\nu_\beta)$ for
  all flavors $\beta=e, \mu$, and $\tau$ are the same.
\begin{align}
  &{\ }&\hspace*{-40mm}
  \epsilon_{NC}(E_\nu,\nu_e)=\epsilon_{NC}(E_\nu,\nu_\mu)=\epsilon_{NC}(E_\nu,\nu_\tau)
\end{align}

\item The detection efficiency for
hadronic shower events from $\nu_\tau$ and $\bar{\nu}_\tau$
is the same as that for electromagnetic shower
events from $\nu_e$ and $\bar{\nu}_e$.
Since 65\% (18\%) of tau decay is hadronic
(electromagnetic)\,\cite{ParticleDataGroup:2020ssz},
this assumption implies
\begin{align}
  &{\ }&\hspace*{-50mm}
  \epsilon_{CC}(E_\nu,\nu_\tau)
  =0.83\,\epsilon_{CC}(E_\nu,\nu_e)\,.
\end{align}
\end{enumerate}

\begin{figure}[H]
  %\begin{center}
    \hspace{1.3cm}
\includegraphics[width=16cm]{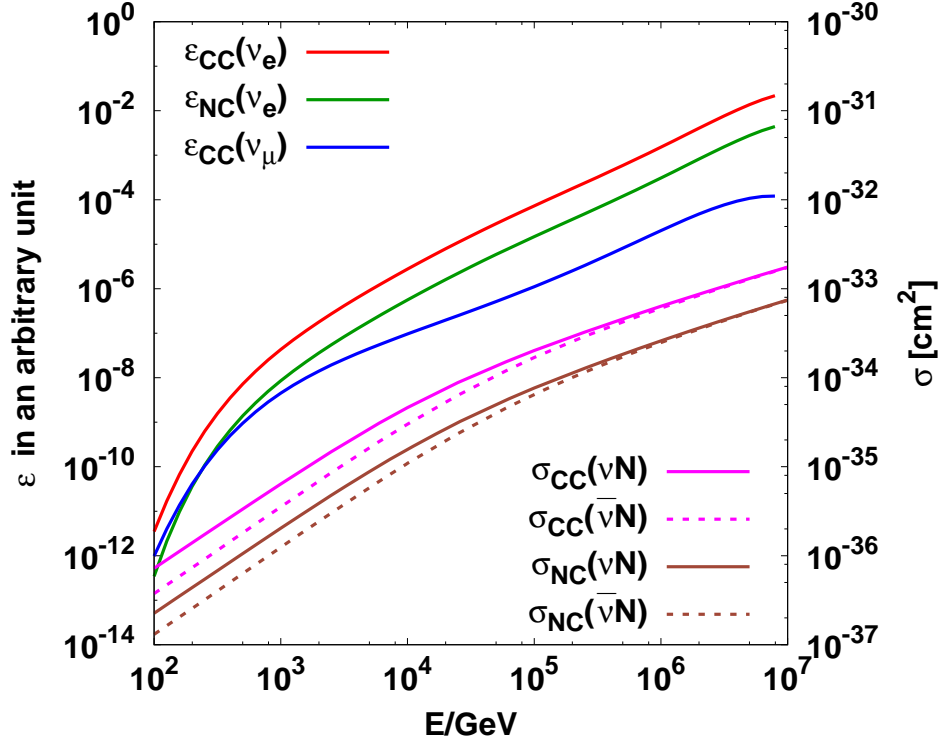}
\caption{\label{fig3} %\it Detection efficiency
  $\epsilon_{CC}(E,\nu_e)$, $\epsilon_{NC}(E,\nu_e)$ and
  $\epsilon_{CC}(E,\nu_\mu)$, which are deduced from FIG.6 in
  Ref.\,\cite{IceCube:2015mgt}, the flux in Refs.\,\cite{Honda:2006qj}
  and \cite{Chirkin:2004ic}, the cross sections in
  Ref.\,\cite{Gandhi:1998ri} {\color{black}with the CTEQ4–DIS parton
    distributions}, and assumptions (i) - (iii) in this article.  The
  absolute magnitude of the detection efficiency is arbitrary, and we
  fix the normalization so that the total numbers of events coincide
  with those in Ref.\,\cite{IceCube:2015mgt}.  {\color{black}The cross
    sections for the charged (neutral) current neutrino-nucleon
    (antineutrino-nucleon) interactions in Ref.\,\cite{Gandhi:1998ri}
    are also shown in the figure.}}
\end{figure}

From TABLE I in Ref.\,\cite{IceCube:2015mgt},
the numbers of events for 332 days at IceCube
are 198 (CC) + 17 (NC) for $\nu_e+\bar{\nu}_e$,
387 (CC) + 258 (NC) for $\nu_\mu+\bar{\nu}_\mu$,
115 for cosmic ray muons, and
103 for astrophysical $\nu_\beta+\bar{\nu}_\beta~(\beta=e+\mu+\tau)$.
Since we know the flux of conventional atmospheric
neutrinos and the cross section at each energy,
the CC and NC numbers of events for
$\nu_e+\bar{\nu}_e$ along with
assumptions (i) and (ii) allow us to deduce
the detection efficiency for 
$\epsilon_{CC}(E,\nu_e)$ and $\epsilon_{NC}(E,\nu_e)$,
where they are extrapolated up to the
neutrino energy $E\lesssim$ 10 PeV.
Then, from the CC and NC numbers of events for
$\nu_\mu+\bar{\nu}_\mu$ with assumptions (iii),
we can deduce $\epsilon_{CC}(E,\nu_\mu)$.
Thus, we obtain $\epsilon_{CC}(E,\nu_e)$, $\epsilon_{NC}(E,\nu_e)$
and $\epsilon_{CC}(E,\nu_\mu)$,
and they are depicted
in an arbitrary unit in Figure \ref{fig3}.
Assumption (iv) can be used to deduce
$\epsilon_{CC}(E,\nu_\tau)$ if there are
$\nu_\tau+\bar{\nu}_\tau$.  

\subsubsection{Astrophysical neutrinos}

The flux of astrophysical neutrinos is assumed to be isotropical, and
the ratios 
$\Phi^{\mbox{\scriptsize ast}}(\nu_\mu)/\Phi^{\mbox{\scriptsize ast}}(\nu_e)$ and
$\Phi^{\mbox{\scriptsize ast}}(\bar{\nu}_\mu)/
\Phi^{\mbox{\scriptsize ast}}(\bar{\nu}_e)$
at the production point are assumed to be 2:
\begin{align}
  &{\ }\Phi^{\mbox{\scriptsize ast}}(\nu_\mu, E, \Theta)
  = \Phi^{\mbox{\scriptsize ast}}(\bar{\nu}_\mu, E, \Theta)
= 2\,\Phi^{\mbox{\scriptsize ast}}(\nu_e, E, \Theta)
  = 2\,\Phi^{\mbox{\scriptsize ast}}(\bar{\nu}_e, E, \Theta)
      \nonumber\\
      =&\,\frac{1}{3}\,
      \Phi^{\mbox{\scriptsize ast}}_0\left(\frac{E}{100 \mathrm{TeV}}\right)^{-\gamma}
\end{align}
where $\gamma$ is the astrophysical spectral index, and
$\Phi^{\mbox{\scriptsize ast}}_0$ is the total flux of astrophysical neutrinos
at $E=$100 TeV.\footnote{
If there is no enhancement of the
sterile mixing angle $\theta_{14}$
due to the matter effect in the Earth, then
each flux at $E=$100 TeV is
$\Phi^{\mbox{\scriptsize ast}}(\nu_e)\simeq
\Phi^{\mbox{\scriptsize ast}}(\bar{\nu}_e)\simeq
\Phi^{\mbox{\scriptsize ast}}(\nu_\mu)\simeq
\Phi^{\mbox{\scriptsize ast}}(\bar{\nu}_\mu)\simeq
\Phi^{\mbox{\scriptsize ast}}(\nu_\tau)\simeq
\Phi^{\mbox{\scriptsize ast}}(\bar{\nu}_\tau)\simeq\Phi^{\mbox{\scriptsize ast}}_0/6$,
so that the total flux is approximately $\Phi^{\mbox{\scriptsize ast}}_0$.
}
We adopt the reference value $\gamma=2.4$ used in
Ref.\,\cite{IceCube:2015mgt}.

If there is no enhancement of the
sterile mixing angle $\theta_{14}$
due to the matter effect in the Earth,
the oscillation probabilities are
approximately
given by those in vacuum in the limit of
an infinite baseline length, and
\begin{align}
{\ }&\sum_{\alpha=e,\mu} \,
\Phi^{\mbox{\scriptsize ast}}(\nu_\alpha, E', \Theta')\,
P(\nu_\alpha\to\nu_e)
      \nonumber\\
\simeq
&\sum_{\alpha=e,\mu} \,
\Phi^{\mbox{\scriptsize ast}}(\nu_\alpha, E', \Theta')\,
P(\nu_\alpha\to\nu_\mu)
      \nonumber\\
\simeq
&\sum_{\alpha=e,\mu} \,
\Phi^{\mbox{\scriptsize ast}}(\nu_\alpha, E', \Theta')\,
P(\nu_\alpha\to\nu_\tau)\,,
\label{flavorratio1}
\end{align}
as was pointed out in Ref.\,\cite{Learned:1994wg}.
Meanwhile, if
there is an enhancement of the
sterile mixing angle $\theta_{14}$,
Eq.\,(\ref{flavorratio1})
is modified for the energy range
in which the enhancement of the
effective mixing angle $\theta_{14}$
occurs.  Particularly, if the
sterile mixing angle $\theta_{14}$
and mass-squared difference
lie in a suitable range, then
the flavor ratio
$(\nu_e+\bar{\nu}_e)$ : $(\nu_\mu+\bar{\nu}_\mu)$ :
$(\nu_\tau+\bar{\nu}_\tau)$
$= 1 - |T_{14}|^2/2 : 1 : 1$,
(where the flavor ratio of neutrinos
is given by Eq.\,(\ref{ratio1}), and the flavor ratio of antineutrinos
is 1:1:1 in the absence of resonance),
may be observed, instead of
the standard flavor ratio\,\cite{Learned:1994wg}
$(\nu_e+\bar{\nu}_e)$ : $(\nu_\mu+\bar{\nu}_\mu)$ : $(\nu_\tau+\bar{\nu}_\tau)$
=1:1:1.
This deviation depends on the neutrino energy,
and it can be a signal to search for
sterile neutrino oscillations.
Although we discuss only the case
with $\theta_{14} \ne 0$, $\theta_{24} = \theta_{34} = 0$
in this study,
if the astrophysical neutrino events
can be separated from those of
atmospheric neutrinos,
the energy dependence of the flavor ratio is
in general
one of the signatures of sterile neutrino
oscillations for the range
$1~\mathrm{eV}^2\lesssim \Delta m^2_{41}\lesssim 100~\mathrm{eV}^2$.

\subsection{$\chi^2$}
\label{Subsec4.2}
In this study, we examine whether
the numbers of shower events
predicted by the (3+1)-scheme
with $\theta_{14}\ne 0$ can be
distinguished from those without
sterile neutrino oscillation,
i.e., those with the standard three-flavor
framework with the central value of
the oscillation parameters given in
Eqs.\,(\ref{osc-param1})-(\ref{osc-param7}) with NO.
To evaluate the significance
between the number of events with $\theta_{14}\ne 0$
and the one without non-zero $\theta_{14}$,
we use the following Poisson likelihood
chi-square:
\begin{align}
  {\ }&\hspace*{-30mm}
\chi^2_{0}
=
2 \sum\limits_{j, k}
\left[ 
Y_{j k}
-
N_{j k}
-
N_{j k} \log\left( \frac{Y_{j k}}{N_{j k}} \right)
\right] \,,
\label{Chi0^2}
\end{align}
where $j$ represents the $j$-th zenith angle bin
($(j-11)/10\le \cos\Theta \le (j-10)/10$,
$1 \le j \le 20$),
and $k$ represents the $k$-th energy bin
($({\color{black}k}+7)/4\le \log_{10}(E/\mathrm{GeV})\le ({\color{black}k}+8)/4$,
$1 \le k \le 16$).  Moreover, 
\begin{align}
Y_{j k}
= &{}
(1+\alpha_1)
\left[ N^{\mbox{\scriptsize atm}}_{j k}( \nu_e + \bar\nu_e; \theta_{14})
+
N^{\mbox{\scriptsize atm}}_{j k}( \nu_\mu + \bar\nu_\mu; 0) \right] \notag \\
 &+
(1+\alpha_2) 
 N^{\mbox{\scriptsize cr}-\mu}_{j k}
 %\notag \\
  %&{\ }&\hspace*{-0mm}
 +
(1+\alpha_3)
N^{\mbox{\scriptsize ast}}_{j k}(\theta_{14})  
\end{align}
and
\begin{align}
\hspace*{-10mm}N_{j k}
=& 
N^{\mbox{\scriptsize atm}}_{j k}( \nu_e + \bar\nu_e; 0)
+
N^{\mbox{\scriptsize atm}}_{j k}( \nu_\mu + \bar\nu_\mu; 0) \notag \\
\hspace*{-10mm}&+
N^{\mbox{\scriptsize cr}-\mu}_{j k}
+
N^{\mbox{\scriptsize ast}}_{j k}(0) \,,
\end{align}
where $\alpha_1$, $\alpha_2$ and $\alpha_3$ are three pull
parameters; $N^{\mbox{\scriptsize atm}}_{j k}( \nu_e +
\bar\nu_e; \theta_{14})$ and $N^{\mbox{\scriptsize atm}}_{j k}( \nu_e +
\bar\nu_e; 0)$ represent the number of events of atmospheric
electron neutrinos with and without {\it sterile} neutrino oscillation
(i.e., $\theta_{14}\ne 0$ or $\theta_{14}= 0$), respectively, while
assuming all three-flavor oscillation parameters
as the central value given in
Eqs.\,(\ref{osc-param1})-(\ref{osc-param7}) with NO
in both cases;
$N^{\mbox{\scriptsize atm}}_{j k}( \nu_\mu + \bar\nu_\mu; 0)$
represents the number of events of atmospheric muon neutrinos without
{\it sterile} neutrino oscillation, $N^{\mbox{\scriptsize ast}}_{j
  k}(\theta_{14})$ and $N^{\mbox{\scriptsize ast}}_{j k}(0)$ represent
the number of events of astrophysical neutrinos with and without {\it
  sterile} neutrino oscillation, respectively, $N^{\mbox{\scriptsize
    cr}-\mu}_{j k}$ stands for the number of events of cosmic ray
muons.
Notice that in all numbers of events
$N^\ast_{jk}(\ast, \theta_{14})$ and
$N^\ast_{jk}(\ast, 0)$, the effects of
the standard three-flavor oscillation
are considered, since we are
interested in the difference between
the numbers of shower events with
the standard three-flavor oscillation scenario
and those with the (3+1)-scheme.

The total $\chi^2$ is defined as follows:
\begin{align}
&{\ }&\hspace*{-40mm}
\chi^2
=\min_{\alpha_1, \alpha_2, \alpha_3}\left[
\chi_{0}^2
+
\left( \frac{\alpha_1}{\sigma_1} \right)^2
+
\left( \frac{\alpha_2}{\sigma_2} \right)^2
+
\left( \frac{\alpha_3}{\sigma_3} \right)^2 \right] \,,
\label{Chi}
\end{align}
where $\sigma_1$, $\sigma_2$ and $\sigma_3$ are the
three systematic errors for atmospheric neutrinos,
cosmic ray muons, and astrophysical neutrinos, respectively.
In this study, we set $\sigma_1 =
\sigma_2 = \sigma_3 = 0.4$ for simplicity.\,\footnote{
  Although various systematic errors
  are given in Ref.\,\cite{IceCube:2020tka}, in our analysis,
  for simplicity, we
  consider only the overall normalization for the atmospheric
  neutrinos ($\sigma_1$), astrophysical neutrinos
  ($\sigma_3$), and cosmic ray muons ($\sigma_2$).
  The reference values quoted in Ref.\,\cite{IceCube:2020tka}
  are $\sigma_1=0.4$ and $\sigma_3=0.36$, but
  we could not find the one for $\sigma_2$.
  For simplicity, we assumed that all of them are 0.4.
Implications of this choice will be discussed in Section \ref{Sec5}.
}
In (\ref{Chi}) we fix the three-flavor oscillation parameters
and assume the central value given in
Eqs.\,(\ref{osc-param1})-(\ref{osc-param7}) with Normal Ordering.
The reason that we do not marginalize $\chi^2$
with respect to the three-flavor oscillation parameters
is that we are interested in fitting the energy
spectrum with a dip (i.e., the one with non-zero $\theta_{14}$)
to that without a dip, and variations with respect to
the three-flavor oscillation parameters are not expected
to affect this fitting.
Thus, $\chi^2$ becomes a function of $\theta_{14}$
and $\Delta m^2_{41}$.

\subsection{Sensitivity}
\label{Subsec4.3}

\begin{figure}[H]
\begin{center}%\vspace{-0.7cm}
\includegraphics[width=12cm]{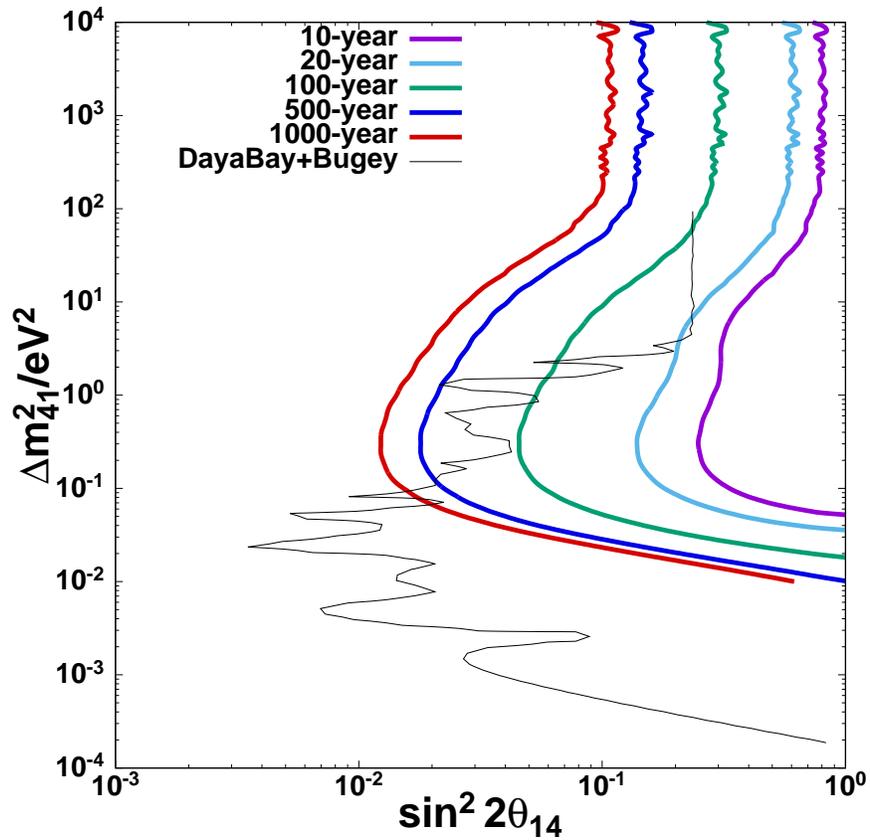}
\caption{\label{fig1} %\it 
  The expected excluded region from the shower events
  of the IceCube experiment for data size
  10, 20, 100, 500, and 1000 years.
  The bound at 90\% CL from the combined analysis of the Bugey3 and
  Daya Bay experiments\,\cite{MINOS:2020iqj} is also shown.}
\end{center}
\end{figure}

\begin{figure}[H]
  %\begin{center}%
  \hspace{1.4cm}
\includegraphics[width=15cm]{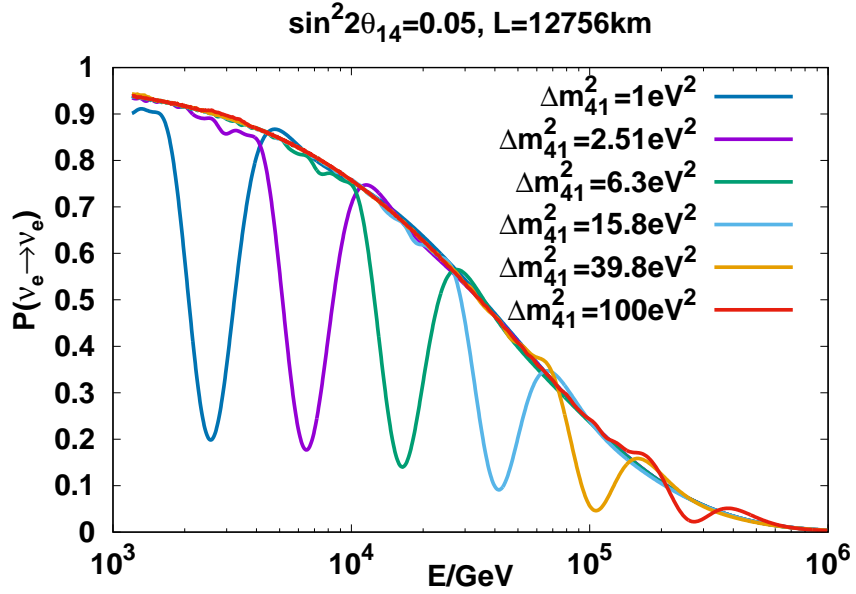}
\caption{\label{fig0} %\it 
  Oscillation probability
  $P(\nu_e\to\nu_e)$ as a function
  of the neutrino energy $E$ for various values
  of $\Delta m^2_{41}$ in the case
  of $\sin^22\theta_{14}=0.05$ and the
  baseline length $L$ = 12756 km (= the diameter
  of the Earth).  For $E\gtrsim$ 100 TeV,
  the probability decreases because of the
  absorption effect of neutrino.
}
\end{figure}

We use the number of events mentioned in (\ref{Subsec4.1}) and
evaluate $\chi^2$ in Eq.\,(\ref{Chi}).
The region with $\chi^2>4.61$ represents the
region in which the hypothesis with
$\theta_{14}\ne 0$ can be excluded
at 90\%CL, and it is depicted
in the ($\theta_{14}$, $\Delta{m}^2_{41}$)-plane in Figure \ref{fig1}.
From Figure \ref{fig1}, if we use the number of events observed
for more than 100 years with the present IceCube detector,
the sensitivity can be better than the one given
by DayaBay+Bugey at 90\% CL\,\cite{MINOS:2020iqj}
for $1~\mathrm{eV}^2\lesssim\Delta{m}^2_{41}\lesssim 100~\mathrm{eV}^2$.
To achieve this sensitivity in the near
future, we must increase the volume of the detector to get more numbers
of events.
We also observe from Figure \ref{fig1} that
the sensitivity to $\Delta{m}^2_{41}$ is saturated
for $\Delta{m}^2_{41} \gtrsim$ 100 eV$^2$.
Figure \ref{fig0} shows the dependence of
the oscillation probability $P(\nu_e\to\nu_e)$,
which is calculated numerically by
considering the absorption effect $B$ in
Eq.\,(\ref{DiracEq}),
as a function of the neutrino energy $E$.
To see the dip, realized
by the neutrino energy
$E_{\mbox{\rm\scriptsize dip}}=\Delta{m}^2_{41}/\{2(A_e-A_s)\}$,
$E_{\mbox{\rm\scriptsize dip}}\lesssim$ 500 TeV
must be satisfied, and this condition
yields $\Delta{m}^2_{41}\lesssim$ 100 eV$^2$.

\begin{figure}[H]
\begin{center}%\vspace{-0.7cm}
\includegraphics[width=12cm]{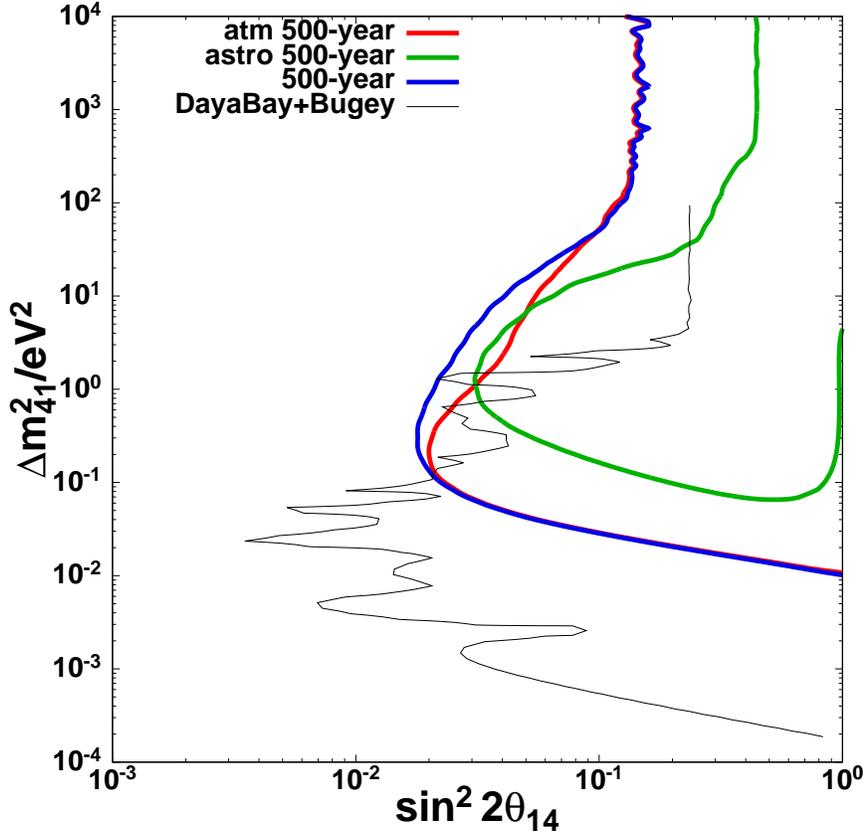}
\caption{\label{fig2} %\it 
  Hypothetical excluded region from the shower events
  of the IceCube experiment for 500 years, assuming
  only atmospheric neutrinos (red curve)
  or only astrophysical neutrinos (green curve),
  where the thin blue curve stands for the region
  with both atmospheric and astrophysical neutrinos for 500 years.
  The bound from the combined analysis of the Bugey3 and
  the Daya Bay experiments\,\cite{MINOS:2020iqj} is also shown
  (the thin black curve).}
\end{center}
\end{figure}

Figure \ref{fig2} shows the contribution of atmospheric and
astrophysical neutrinos to the
sensitivity expected by
observation for 500 years at the present IceCube.
We observe that atmospheric neutrinos contribute more to the
sensitivity than astrophysical neutrinos
because the number of events of the former is
larger.
Moreover,
the peak of the sensitivity is obtained
at $\Delta{m}^2_{41}\sim$ 0.1 eV$^2$
($\Delta{m}^2_{41}\sim$ 1 eV$^2$)
for atmospheric (astrophysical) neutrinos,
which is because the number of
events of atmospheric neutrinos
(astrophysical neutrinos) has a peak
at $E\sim$ 1 TeV ($E\sim$ 10 TeV)
in the energy spectrum (cf. FIG.6 in Ref.\,\cite{IceCube:2015mgt}),
and the condition for $\Delta{m}^2_{41}$
to have the dip
(the denominator of Eq.\,(\ref{14.1}) vanishes)
leads to $\Delta{m}^2_{41}\sim$ 0.1 eV$^2$
($\Delta{m}^2_{41}\sim$ 1 eV$^2$).

We note in passing that the curve
from astrophysical neutrinos in Figure \ref{fig2}
has two intersections with
a straight line $\Delta{m}^2_{41}$=constant
for $0.06~\mbox{\rm eV}^2\lesssim
\Delta{m}^2_{41}\lesssim O(1)$ eV$^2$,
i.e., there are two solutions
of $\chi^2=4.61$ for a fixed value of
$\Delta{m}^2_{41}$ in this region of $\Delta{m}^2_{41}$,
which can be qualitatively understood as follows.
As described in Appendix \ref{appendix-e},
as $\theta_{14}\to\pi/4$, if
$\Delta E \equiv\Delta{m}^2_{41}/(2E) \ll |A_e-A_s|$,
the $\nu_e$ flux becomes approximately independent
of the neutrino energy $E$ and baseline length
$L$ in the Earth.  In this case, if the systematic error
is not very small, the suppression due to neutrino oscillation
is constant and the data may be fit by
considering the overall normalization uncertainty.
Meanwhile, if $\Delta{m}^2_{41}$ becomes
large, the condition
$\Delta{m}^2_{41}/(2E) \ll |A_e-A_s|$
is no longer satisfied, and the above argument
does not hold.  This argument qualitatively explains
the behavior of the curve
from astrophysical neutrinos in Figure \ref{fig2}.

\section{Conclusions}
\label{Sec5}

In this study we showed that a future neutrino
telescope facility, whose volume is at least
10 times as large as that of IceCube,
can improve sensitivity to $\theta_{14}$
by running for 10 years
for 1 eV$^2 \lesssim \Delta m^2_{41}\lesssim$100 eV$^2$
in the (3+1)-scheme{\color{black}, if no dip is observed
in the energy spectrum of the shower events}.

We have assumed that $\theta_{24}=\theta_{34}=0$
in our numerical analysis.
If we turn on small $\theta_{24}$ and
$\theta_{34}$, the resonance in the neutrino mode has small
perturbation with respect to $\theta_{24}$ and $\theta_{34}$, and we
expect that our conclusion do not change very much.

Although the variety of systematic errors
in this study
is smaller than that in Ref.\,\cite{IceCube:2020tka},
it is not expected to significantly affect
the sensitivity to $\sin^22\theta_{14}$ at
the peak, since the sensitivity
is mainly given by the energy
spectrum's shape.  What is significantly affected by the details
of the systematic errors is the behavior for
$\Delta m^2_{41}\gtrsim$ 100 eV$^2$, because
the oscillation probability in the region
is expected to be averaged out to constant,
and the details of the overall normalization
significantly differ.
So, the sensitivity to $\sin^22\theta_{14}$
for $\Delta m^2_{41}\gtrsim$100 eV$^2$
is expected to depend on the details
of the systematic errors, and more
careful analysis would be required.

In our analysis, we assumed that there is
no prompt neutrino event.
If the volume of the future detector is
increased, there can be a small contribution
of prompt neutrino events for
the neutrino energy $E\gtrsim$ 10$^{4.5}$ GeV.
However, from the result of Ref.\,\cite{IceCube:2015mgt}
in which the data agree with a hypothesis
of zero prompt neutrino event, 
the contribution from prompt neutrinos is
expected to be much smaller than that of
astrophysical neutrinos.
Therefore, we expect that our results are not significantly
altered even if the contribution from prompt neutrinos is
considered.

We studied only the shower events for simplicity.  If we
combine the track and shower events, we can even give
stronger constraints on the sterile mixing angles.  In the presence of
all non-zero sterile mixing angles, it is expected that we can also see
a dip in the antineutrino disappearance channel
$\bar\nu_\mu\to\bar\nu_\mu$ at the same energy region as that in
neutrino disappearance channel $\nu_e\to\nu_e$.  This dip can be
observed in the track events, and it gives us a constraint on
$\varphi_{13}=\sin^{-1}(c_{24}c_{34})$, so we could
obtain more information on the sterile mixing angles than
in this study.

{\color{black}In this article we obtain the sensitivity to $\theta_{14}$
on the assumption that no dip is observed.
If the dip is discovered, however, the situation becomes
different from that of the present work.  By looking at the shower events only,
it is impossible to conclude whether the dip comes
from the neutrino mode which is caused mainly by
the enhancement due to $\theta_{14}$,
from the antineutrino mode which is caused mainly by
the enhancement due to $\theta_{24}$ and $\theta_{34}$, or
from both in the presence of all $\theta_{j4}~(j=1,2,3)$.
To solve such degeneracy, we need information
from the track events.  Discussions on such degeneracy
is complicated, and it will be a future work.}

In summary, we emphasize that measurements of high energy
neutrinos at a neutrino telescope have a potential advantage over
low energy short baseline experiments in investigating
sterile neutrino oscillations for 1 eV$^2~\lesssim \Delta
m_{41}^2\lesssim$ 100 eV$^2$, if the volume of the detector
is at least 10 times as large as that of IceCube.

\section*{Acknowledgements}
This research was partly supported by a Grant-in-Aid for Scientific
Research of the Ministry of Education, Science and Culture, under
Grants No. 18K03653, No. 18H05543 and No. 21K03578.

\newpage
\vglue 10mm
\noindent
{\Large\bf Appendix}

\appendix

\section{Derivation of Eq.\,(\ref{P1}) in the case of $\theta_{14}\ne 0$,
and $\theta_{24} = \theta_{34} = 0$}
\label{appendix-a}
In this Appendix, we discuss oscillations of atmospheric neutrinos
for $\theta_{14} \ne 0$, and $\theta_{24} = \theta_{34} = 0$.
Since $\nu_\mu$ and $\nu_\tau$ are decoupled
from $\nu_e$ and $\nu_s$, the system can
be effectively described by the two-flavor framework.
The only remaining CP phase $\delta_2$
in Eq.\,(\ref{u}) disappears by absorbing
in the flavor eigenstate $\nu_\alpha~(\alpha=e,s)$.
Thus, we express the evolution equation as follows:
\begin{align}
i \frac{d}{dt}
\left( \begin{array}{rl}
\nu_e \\
\nu_s 
\end{array} \right)
=
\left[
U_1
\left( \begin{array}{cc}
0 & 0             \\   
0 & \Delta E \\   
\end{array} \right)
U_1^{-1}
+
\left( \begin{array}{cc}
A_e & 0        \\   
0        & A_s \\  
\end{array} \right)
\right]
\left( \begin{array}{rl}
\nu_e     \\
\nu_s 
\end{array} \right) \,,
\label{DiracEq.1}
\end{align}
where {\color{black}$\Delta E \equiv \sqrt{p^2+m^2_2}-\sqrt{p^2+m^2_1}\simeq
(m_2^2-m_1^2)/(2p) \equiv \Delta m^2/(2p) \simeq \Delta m^2/(2E)$,}
\begin{eqnarray}
  &{\ }&\hspace*{-80mm}
  U_1 = \exp(i \theta_{14} \sigma_2)\,,
\end{eqnarray}
and $\sigma_2$ is one of the three Pauli matrices
\begin{align}
  \hspace*{-5mm}
  \sigma_1=
\left( \begin{array}{cc}
0 & 1 \\
1 & 0 \\
\end{array} \right) \,,~~
\sigma_2=
\left( \begin{array}{cc}
0 & -i  \\
i  & 0  \\
\end{array} \right) \,,~~
\sigma_3=
\left( \begin{array}{cc}
1 & 0  \\
0 & -1 \\
\end{array} \right) \,.
\end{align}
We denote the Hamiltonian of Eq.\,(\ref{DiracEq.1}) as $H_1$
and rewrite it as follows:
\begin{align}
H_1
& =
\exp(i \theta_{14} \sigma_2)
\frac{\Delta E}{2} ({\bf 1}_2 - \sigma_3)
\exp(-i \theta_{14} \sigma_2)
+
\frac{A_e}{2} ({\bf 1}_2 + \sigma_3)
+
\frac{A_s}{2} ({\bf 1}_2 - \sigma_3) \notag \\
& =
\frac{\Delta E + A_e + A_s}{2} {\bf 1}_2
-\frac{1}{2}
(\alpha \sigma_3 - \beta \sigma_1) \,,
\label{H1.1}
\end{align}
where
\begin{align} 
  \hspace*{-20mm}
\alpha 
= 
\Delta E \cos{2\theta_{14}} - A_e + A_s, 
\ 
\beta = \Delta E \sin{2\theta_{14}} \,,
\end{align}
and ${\bf 1}_2$ is the $2\times 2$ identity matrix.
Denoting the mixing
angle in matter as $\tilde\theta_{14}$, we have
\begin{align}  
&\exp(-i \tilde\theta_{14} \sigma_2)
(\alpha \sigma_3 - \beta \sigma_1)
\exp(i \tilde\theta_{14} \sigma_2) 
\notag \\
=&
(\alpha \cos{2\tilde\theta_{14}} + \beta \sin{2\tilde\theta_{14}}) \sigma_3
 +
(\alpha \sin{2\tilde\theta_{14}} - \beta \cos{2\tilde\theta_{14}}) \sigma_1 \,.
\end{align}
If we define 
\begin{align}
  \hspace*{-30mm}
\tan{2\tilde\theta_{14}} 
= 
\frac{\beta}{\alpha}
= 
\frac{\Delta E \sin{2\theta_{14}}}{\Delta E \cos{2\theta_{14}} - A_e + A_s} \,,
\label{14.1}
\end{align}
which implies
\begin{align}
  \hspace*{-20mm}
\sin{2\tilde\theta_{14}} 
= 
\frac{\beta}{\sqrt{\alpha^2 + \beta^2}} \,,
\ 
\cos{2\tilde\theta_{14}} 
= 
\frac{\alpha}{\sqrt{\alpha^2 + \beta^2}} \,,
\end{align}
we obtain
\begin{align}
  \hspace*{-10mm}
\alpha \sigma_3 - \beta \sigma_1
=
\sqrt{\alpha^2 + \beta^2}
\exp(i \tilde\theta_{14} \sigma_2)
\sigma_3
\exp(-i \tilde\theta_{14} \sigma_2) \,.
\end{align}
Further, if we define
\begin{align}
\Delta\tilde{E}_1 
= 
\sqrt{\alpha^2 + \beta^2} 
=
\sqrt{(\Delta E \cos{2\theta_{14}} - A_e + A_s)^2 + (\Delta E \sin{2\theta_{14}})^2} \,,
\label{E1.1}
\end{align}
Eq.\,(\ref{H1.1}) becomes
\begin{align}
H_1
& =
\exp(i \tilde\theta_{14} \sigma_2)
\left(
\frac{\Delta E + A_e + A_s}{2} {\bf 1}_2
-
\frac{1}{2} \Delta\tilde{E}_1 \sigma_3
\right)
\exp(-i \tilde\theta_{14} \sigma_2) \notag \\
& =
\exp(i \tilde\theta_{14} \sigma_2)
\left( \begin{array}{cc}
\varepsilon_{1 -}  & 0                          \\
0                          & \varepsilon_{1 +}
\end{array} \right)
\exp(-i \tilde\theta_{14} \sigma_2) \,,
\label{H1.2}
\end{align}
where
\begin{align} 
\varepsilon_{1 \mp}
=
\frac{\Delta E + A_e + A_s}{2}
\mp
\frac{1}{2} \Delta\tilde{E}_1 \,.
\label{epsilon1}
\end{align}
From these discussions, the
oscillation probability in this case
is obtained by replacing $\theta_{14}$ by
$\tilde\theta_{14}$ and $\Delta E$
by $\Delta\tilde{E}_1$ in the vacuum
oscillation probability, and we obtain
Eq.\,(\ref{P1}).

\section{Derivation of Eqs.\,(\ref{P3}) and (\ref{P4})
  in the case of $\theta_{14} = 0$, $\theta_{24}\ne 0$, and $\theta_{34}\ne 0$}
\label{appendix-b}

In this Appendix, we discuss oscillations of atmospheric neutrinos
for $\theta_{14} = 0$, $\theta_{24} \ne 0$, and $\theta_{34} \ne
0$.  In this case, it turns out that resonance occurs in the
antineutrino mode.  So, we start with Eq.\,(\ref{DiracEq-anti}) for
antineutrinos.  There is mixing among $\bar\nu_\mu$,
$\bar\nu_\tau$ and $\bar\nu_s$, and Eq.\,(\ref{DiracEq-anti})
reduces to that in the three-flavor case
\begin{align}
i \frac{d}{dt}
\left( \begin{array}{rl}
\bar\nu_\mu \\ 
\bar\nu_\tau \\
\bar\nu_s
\end{array} \right)
=
\left[
U_2^\ast
\left( \begin{array}{ccc}
0 & 0 & 0             \\   
0 & 0 & 0             \\   
0 & 0 & \Delta E \\   
\end{array} \right)
U_2^{\ast-1}
+
\left( \begin{array}{ccc}
0 & 0 & 0        \\   
0 & 0 & 0        \\   
0 & 0 & -A_s \\  
\end{array} \right)
\right]
\left( \begin{array}{rl}
\bar\nu_\mu \\ 
\bar\nu_\tau \\
\bar\nu_s
\end{array} \right) \,. 
\label{DiracEq.2}
\end{align}
In the discussions of the three-flavor mixing case,
it is useful to introduce the following
three Gell-Mann matrices:
\begin{align}
\lambda_2= 
\left( \begin{array}{ccc}
0 & -i & 0 \\
i  & 0 & 0 \\   
0 & 0 & 0     
\end{array} \right) \,,~~
\lambda_5= 
\left( \begin{array}{ccc}
0 & 0 & -i \\
0 & 0 & 0 \\   
i  & 0  & 0     
\end{array} \right) \,,~~
\lambda_7= 
\left( \begin{array}{ccc}
0 & 0 & 0 \\
0 & 0 & -i \\   
0 & i  & 0     
\end{array} \right)\,.
\end{align}
Now using the property
of $3\times 3$ matrices
\begin{align}
  &\mbox{\rm diag}
\left(1, e^{-i\delta_3/2}, e^{i\delta_3/2}\right)
=\mbox{\rm diag}
\left(1, e^{i\delta_3/2}, e^{i\delta_3/2}\right)
\cdot\mbox{\rm diag}\left(1, e^{-i\delta_3}, 1\right),
\nonumber\\
&\exp(i \theta_{34} \lambda_7) \cdot
\mbox{\rm diag}\left(1, e^{i\delta_3/2},
e^{i\delta_3/2}\right)
=\mbox{\rm diag}\left(1, e^{i\delta_3/2},
e^{i\delta_3/2}\right)\cdot\exp(i \theta_{34} \lambda_7)\,,
\nonumber\\
&\mbox{\rm diag}\left(1, e^{-i\delta_3}, 1\right)\cdot
\exp(i \theta_{24} \lambda_5)
=\exp(i \theta_{24} \lambda_5)\cdot
\mbox{\rm diag}\left(1, e^{-i\delta_3}, 1\right)\,,
\nonumber
\end{align}
in the limit of one mass scale dominance,
we can show that the diagonal matrix
$\mbox{\rm diag}(1, e^{i\delta_3/2},
e^{i\delta_3/2})$
($\mbox{\rm diag}(1, e^{-i\delta_3}, 1)$)
with a CP phase $\delta_3$ can be moved
to the left of $\exp(i \theta_{34} \lambda_7)$
(the right of $\exp(i \theta_{24} \lambda_5)$),
respectively:
\begin{align}
U_2^\ast &= \mbox{\rm diag}
\left(1, e^{i\delta_3/2}, e^{-i\delta_3/2}\right)\cdot
\exp\left(i \theta_{34} \lambda_7\right) \cdot
\mbox{\rm diag}\left(1, e^{-i\delta_3/2}, e^{i\delta_3/2}\right)\cdot
\exp\left(i \theta_{24} \lambda_5\right)
\nonumber\\
&=\mbox{\rm diag}
\left(1, e^{i\delta_3/2}, e^{-i\delta_3/2}\right)\cdot
\mbox{\rm diag}\left(1, e^{i\delta_3/2},
e^{i\delta_3/2}\right)\cdot
\exp\left(i \theta_{34} \lambda_7\right)
\nonumber\\
&\quad\times\exp\left(i \theta_{24} \lambda_5\right)\cdot
\mbox{\rm diag}\left(1, e^{-i\delta_3}, 1\right)
\nonumber\\
&=\mbox{\rm diag}
\left(1, e^{i\delta_3}, 1\right)\cdot
\exp\left(i \theta_{34} \lambda_7\right)\cdot
\exp\left(i \theta_{24} \lambda_5\right)\cdot
\mbox{\rm diag}\left(1, e^{-i\delta_3}, 1\right)\,.
\end{align}
The phase factor
$\mbox{\rm diag}(1, e^{-i\delta_3}, 1)$ cancels
in $U_2^\ast\,\mbox{\rm diag}(0,0,\Delta E)\,U_2^{\ast-1}$,
and by redefining $(\bar\nu_\mu, \bar\nu_\tau, \bar\nu_s)^T$
by $\mbox{\rm diag}(1, e^{-i\delta_3}, 1)
(\bar\nu_\mu, \bar\nu_\tau, \bar\nu_s)^T$,
the CP phase $\delta_3$ disappears in the
one mass scale dominance limit.
Therefore, in the following discussions, 
we set $\delta_3=0$ for simplicity.
In this case, there is no difference between
$U_2$ and $U_2^\ast$, so we denote
$U_2^\ast$ as $U_2$ for simplicity
in the following discussions.
To transform Eq.\,(\ref{DiracEq.2})
into a more familiar form,
we introduce the following matrix
\begin{align}
R =
\left( \begin{array}{ccc}
0 & 0 & 1 \\
0 & 1 & 0 \\   
1 & 0 & 0     
\end{array} \right)
\quad
(R^2 = {\bf 1}_3) \,,
\end{align}
where ${\bf 1}_3$ is the $3\times 3$ identity matrix.
Then the Hamiltonian $H_2$ on the right
hand side of Eq.\,(\ref{DiracEq.2})
can be transformed into
\begin{align}
H_2
=
R \left[
R U_2 
\left( \begin{array}{ccc}
0 & 0 & 0             \\   
0 & 0 & 0             \\   
0 & 0 & \Delta E    
\end{array} \right)
U_2^{-1} R
+
\left( \begin{array}{ccc}
-A_s & 0 & 0 \\   
0        & 0 & 0 \\   
0        & 0 & 0    
\end{array} \right)
\right] R \,. 
\label{H2.1}
\end{align}
Next we define 
\begin{align}
H_2^{'}
=
R U_2 
\left( \begin{array}{ccc}
0 & 0 & 0             \\   
0 & 0 & 0             \\   
0 & 0 & \Delta E    
\end{array} \right)
U_2^{-1} R
+
\left( \begin{array}{ccc}
-A_s & 0 & 0 \\   
0        & 0 & 0 \\   
0        & 0 & 0    
\end{array} \right) \,,
\label{H2.2}
\end{align}
and find that $H_2^{'}$ is similar to the Hamiltonian of the evolution
equation of flavor eigenstates of neutrinos in the standard three-flavor
framework including the matter effect, when $\Delta E_{21} =
\Delta{m}^2_{21} / (2E)$ vanishes.  Therefore, in comparison with the
standard PMNS matrix (the CP phase is defined as $\eta$), which is
marked as $U_{PMNS}$ in the following, we have
\begin{align}
R U_2
=
U_{PMNS}
=
\exp(i \varphi_{23} \lambda_7) \cdot
\Gamma(\eta)\cdot
\exp(i \varphi_{13} \lambda_5) \cdot
\Gamma(\eta)^{-1}\cdot
\exp(i \varphi_{12} \lambda_2) \,, 
\label{R U_2}
\end{align}
where 
\begin{eqnarray}
&{\ }&\hspace*{-50mm}
\Gamma(\eta) = \mbox{\rm diag}
\left(e^{-i\eta/2}, 0, e^{i\eta/2}
\right)\,.
\end{eqnarray}
Note that we have used $\varphi_{23}$, $\varphi_{13}$ and
$\varphi_{12}$ to avoid confusion with the original mixing angles
$\theta_{23}$, $\theta_{13}$, and $\theta_{12}$ in the $4\times 4$
matrix $U$ in Eq.\,(\ref{u}).
To make the sign of the first row consistent with the
standard representation $U_{PMNS}$,
we have to introduce a phase matrix, and we have
\begin{align}
R U_2
& =
\left( \begin{array}{ccc}
-s_{24}c_{34} & -s_{34} & c_{24}c_{34} \\
-s_{24}s_{34} &  c_{34} & c_{24}s_{34} \\
c_{24}             & 0           & s_{24}  
\end{array} \right)                           \notag \\                                    
& =
\mbox{\rm diag}
\left(-1, 1, 1\right)\cdot U_{PMNS}
\nonumber\\
& =\left( \begin{array}{ccc}
-C_{12}C_{13} & -S_{12}C_{13} & -e^{-i\eta}\,S_{13} \\
-S_{12}C_{23} - e^{i\eta}\,C_{12}S_{13}S_{23} 
& 
C_{12}C_{23} - e^{i\eta}\,S_{12}S_{13}S_{23} 
&
C_{13}S_{23}                                         \\
S_{12}S_{23} - e^{i\eta}\,C_{12}S_{13}C_{23} 
& 
-C_{12}S_{23} - e^{i\eta}\,S_{12}S_{13}C_{23} 
&
C_{13}C_{23}      
\end{array} \right) \,, 
\label{U_{PMNS}}
\end{align}
where $\mbox{\rm diag}\,(-1, 1, 1)$
is a diagonal matrix with a phase that does not
affect the oscillation probability, and
we have defined
$s_{ij} = \sin{\theta_{ij}}$, 
$c_{ij} = \cos{\theta_{ij}}$, 
$S_{ij} = \sin{\varphi_{ij}}$ 
and
$C_{ij} = \cos{\varphi_{ij}}$.
From Eq.\,(\ref{U_{PMNS}}),
we observe that the CP phase must
satisfy $\eta=\pi$
for variables $S_{ij}$ and $C_{ij}$,
which are semipositive definite, to satisfy
$[R U_2]_{32}=-C_{12}S_{23} - e^{i\eta}\,S_{12}S_{13}C_{23}=0$.
In this case, we have
\begin{eqnarray}
&{\ }&\hspace*{-60mm}
\tan\varphi_{12}=\frac{\tan\varphi_{23}}{S_{13}}\ ;
\end{eqnarray}
from this we can derive
the following three components of $U_{PMNS}$:
\begin{align}
\hspace*{-10mm}
[R U_2]_{21}=&  -S_{12}C_{23} - e^{i\eta}\,C_{12}S_{13}S_{23}
  \nonumber\\
  =&-S_{12}C_{23} +C_{12}S_{13}S_{23}
  \nonumber\\
  =&C_{12}\left(-C_{23}\tan\varphi_{12}+S_{13}S_{23}\right)
  \nonumber\\
  =&C_{12}\left(-C_{23}\frac{\tan\varphi_{23}}{S_{13}}+S_{13}S_{23}\right)
  \nonumber\\
  =&-\frac{C_{12}S_{23}C_{13}^2}{S_{13}}<0\,,
  \\
  \nonumber\\
\hspace*{-10mm}
[R U_2]_{22}=&  C_{12}C_{23} - e^{i\eta}\,S_{12}S_{13}S_{23}
  \nonumber\\
  =&C_{12}C_{23} +S_{12}S_{13}S_{23}
  \nonumber\\
  =&C_{12}\left(C_{23} +S_{13}S_{23}\tan\varphi_{12}\right)
  \nonumber\\
  =&C_{12}\left(C_{23} +S_{13}S_{23}\frac{\tan\varphi_{23}}{S_{13}}\right)
  \nonumber\\
  =&\frac{C_{12}}{C_{23}}
  \left(C_{23}^2 +S_{23}^2\right)
  =\frac{C_{12}}{C_{23}}>0\,,
\\
  \nonumber\\
\hspace*{-10mm}
[R U_2]_{31}=&  S_{12}S_{23} - e^{i\eta}\,C_{12}S_{13}C_{23}
  \nonumber\\
  =&S_{12}S_{23} + C_{12}S_{13}C_{23}
  \nonumber\\
  =&C_{12}\left(S_{23}\tan\varphi_{12} + S_{13}C_{23}\right)
  \nonumber\\
  =&C_{12}\left(S_{23}\frac{\tan\varphi_{23}}{S_{13}} + S_{13}C_{23}\right)
  \nonumber\\
  =&\frac{C_{12}}{C_{23}S_{13}}
  \left(S_{23}^2 + S_{13}^2C_{23}^2\right) >0\,,
\end{align}
and obtain
\begin{align}
&\mbox{\rm diag}
\left(-1, 1, 1\right)\cdot U_{PMNS}
\nonumber\\
 =&\left( \begin{array}{ccc}
-C_{12}C_{13} & -S_{12}C_{13} & S_{13} \\
\displaystyle-\frac{C_{12}S_{23}C_{13}^2}{S_{13}}
& 
\displaystyle\frac{C_{12}}{C_{23}}
&
C_{13}S_{23}                                         \\
\displaystyle\frac{C_{12}}{C_{23}S_{13}}
  \left(S_{23}^2 + S_{13}^2C_{23}^2\right)
& 
0
&
C_{13}C_{23}      
\end{array} \right) \,.
\label{UPMNS2}
\end{align}
From Eqs.\,(\ref{U_{PMNS}}) and (\ref{UPMNS2})
we obtain
\begin{align}
\hspace*{-10mm}
\tan{\varphi_{12}} 
&= 
\frac{t_{34}}{s_{24}} 
\tag{\ref{phi12}}\\
\hspace*{-10mm}
\sin{\varphi_{13}} &= c_{24}c_{34}
\tag{\ref{phi13}}\\
\hspace*{-10mm}
\tan{\varphi_{23}} 
&= 
\frac{s_{34}}{t_{24}} \,,
\tag{\ref{13.1}}
\end{align}
where $t_{ij} = \tan{\theta_{ij}}~(i,j=1,\cdots,4)$.
Eqs.\,(\ref{phi12}), (\ref{phi13}), (\ref{13.1})
are consistent with all elements
of $RU_2$ in Eq.\,(\ref{U_{PMNS}}) for
the region $0\le \theta_{i4}\le \pi/2~(i=2,3)$,
$0\le \varphi_{ij}\le \pi/2~(i,j=1,2,3)$.
Based on Eqs.\,(\ref{H2.1}), (\ref{H2.2}) and (\ref{R U_2}),
we can rewrite the Hamiltonian $H_2$ as follows:
\begin{align}
H_2
= {} &
R \exp(i \varphi_{23} \lambda_7) \notag \\ 
& \times
\left[
\exp(i \varphi_{13} \lambda_5)
\left( \begin{array}{ccc}
0 & 0 & 0             \\   
0 & 0 & 0             \\   
0 & 0 & \Delta E    
\end{array} \right)
\exp(-i \varphi_{13} \lambda_5)
+
\left( \begin{array}{ccc}
-A_s & 0 & 0 \\   
0        & 0 & 0 \\   
0        & 0 & 0    
\end{array} \right) 
\right] \notag \\
& \times
\exp(-i \varphi_{23} \lambda_7) R \,.
\label{H2.3}
\end{align}
Similarly to the case in Subsubsection \ref{Subsubsec3.1.1}, we have
\begin{eqnarray}
&{\ }&\hspace*{-30mm}
H_2
= {} 
R 
\exp(i \varphi_{23} \lambda_7) 
\exp(i \tilde\varphi_{13} \lambda_5) \notag \\
&{\ }&\hspace*{-20mm}
 \times
\left( \begin{array}{ccc}
\varepsilon_{2 -} & 0 & 0 \\   
0                         & 0 & 0 \\   
0 & 0 & \varepsilon_{2 +}    
\end{array} \right) \notag \\
&{\ }&\hspace*{-20mm}
 \times
\exp(-i \tilde\varphi_{13} \lambda_5)
\exp(-i \varphi_{23} \lambda_7) 
R \,.
\label{H2.4}
\end{eqnarray}
where
\begin{align} 
\hspace*{-40mm}
\tan{2\tilde\varphi_{13}}
=
\frac{\Delta E \sin{2\varphi_{13}}}
{\Delta E \cos{2\varphi_{13}} + A_s}\,,
\tag{\ref{13.2}}
\end{align} 
\begin{align}
\hspace*{-40mm}
\varepsilon_{2 \mp}
=
\frac{\Delta E - A_s}{2}
\mp
\frac{1}{2} \Delta\tilde{E}_2 \,,
\tag{\ref{epsilon2.1}}
\end{align}
and
\begin{align}
\Delta\tilde{E}_2 
=
\sqrt{(\Delta E \cos{2\varphi_{13}} + A_s)^2
  + (\Delta E \sin{2\varphi_{13}})^2} \,.
\tag{\ref{E2.1}}
\end{align}

Since the mixing between active and sterile neutrinos
are expected to be very weak in the (3+1)-scheme, we can consider
\begin{eqnarray}
\hspace*{-40mm}
|\theta_{24}| \ll 1, ~~
\ 
|\theta_{34}| \ll 1 \,.
\end{eqnarray}
Then, from (\ref{phi13}) we obtain
\begin{align} 
\hspace*{-30mm}
\varphi_{13} \simeq \frac{\pi}{2} \,.
\hspace*{30mm}
\end{align}
Thus, Eq.\,(\ref{13.2}) becomes
\begin{align} 
\hspace*{-40mm}
\tan{2\tilde\varphi_{13}}
\simeq
\frac{\Delta E \sin{2\varphi_{13}}}
{-\Delta E + A_s} \,.
\label{13.3}
\end{align} 
Eq.\,(\ref{13.3}) indicates that
there is resonance in the disappearance
channel $\bar\nu_\mu\to \bar\nu_\mu$
and the appearance channel
$\bar\nu_\mu\to \bar\nu_\tau$.
From these,
the oscillation probabilities $P(\bar\nu_\mu \to \bar\nu_\mu)$
and $P(\bar\nu_\mu \to \bar\nu_\tau)$ can be, respectively, expressed as follows:
\begin{align}
P(\bar\nu_\mu \to \bar\nu_\mu)
= {} &
1 -
C_{23}^4\sin^22\tilde\varphi_{13}\,
\sin^2\left(\frac{\Delta\tilde{E}_2 L}{2}\right)
\nonumber\\
&~~-\sin^22\varphi_{23}\,
\left\{\tilde{S}_{13}^2
\sin^2\left(\frac{\varepsilon_{2 -} L}{2}\right)
+\tilde{C}_{13}^2 
\sin^2\left(\frac{\varepsilon_{2 +} L}{2}\right)
\right\}\,,
\tag{\ref{P3}}\\
P(\bar\nu_\mu \to \bar\nu_\tau)
={} &
\sin^22\varphi_{23}
  \left\{
  \tilde{S}_{13}^2
\sin^2\left(\frac{\varepsilon_{2 -} L}{2}\right)
+\tilde{C}_{13}^2 
\sin^2\left(\frac{\varepsilon_{2 +} L}{2}\right)
\right.\nonumber\\
&\hspace*{20mm}-\left.\frac{1}{4}
\sin^22\tilde\varphi_{13}\,
\sin^2\left(\frac{\Delta\tilde{E}_2 L}{2}\right)
\right\}\,,
\tag{\ref{P4}}
\end{align}

\section{Derivation of Eq.\,(\ref{ratio1}) in the case of
  $\theta_{14}\ne 0$, and $\theta_{24} = \theta_{34} = 0$}
\label{appendix-c}

In this Appendix, we discuss oscillations of astrophysical neutrinos
for $\theta_{14} \ne 0$, and $\theta_{24} = \theta_{34} = 0$.
As in the discussion in Subsubsection \ref{Subsubsec3.1.1}, we consider
the case of the neutrino mode.  Based on Eq.\,(\ref{H1.2}) in
Subsubsection \ref{Subsubsec3.1.1}, it is not difficult to explicitly obtain the
transition matrix $T$:
\begin{align}
T
&=\exp(i \tilde\theta_{14} {\cal T}_{14})
\cdot
\left( \begin{array}{cccc}
e^{-i L \varepsilon_{1 -}} & 0 & 0 & 0 \\
0                                      & 1 & 0 & 0 \\
0                                      & 0 & 1 & 0 \\
0 & 0 & 0 & e^{-i L \varepsilon_{1 +}}
\end{array} \right)
\cdot
\exp(-i \tilde\theta_{14} {\cal T}_{14})
\notag \\
& =
\left( \begin{array}{cccc}
\tilde{c}_{14}^2 e^{-i L \varepsilon_{1 -}} 
+ 
\tilde{s}_{14}^2 e^{-i L \varepsilon_{1 +}}
& 0 & 0 & 
\tilde{s}_{14} \tilde{c}_{14} 
\left( e^{-i L \varepsilon_{1 +}} 
- e^{-i L \varepsilon_{1 -}} \right)       \\
0                               & 1 & 0 & 0 \\
0                               & 0 & 1 & 0 \\
\tilde{s}_{14} \tilde{c}_{14} 
\left( e^{-i L \varepsilon_{1 +}} 
- e^{-i L \varepsilon_{1 -}} \right) 
& 0 & 0 & 
\tilde{s}_{14}^2 e^{-i L \varepsilon_{1 -}} 
+ 
\tilde{c}_{14}^2 e^{-i L \varepsilon_{1 +}}
\end{array} \right)
\,,
\end{align}
where
\begin{align}
  \hspace*{-40mm}
  {\cal T}_{14}= 
\left( \begin{array}{cccc}
0 & 0 & 0 & -i \\
0 & 0 & 0 & 0 \\
0 & 0 & 0 & 0 \\   
i  & 0 & 0 & 0     
\end{array} \right).
\end{align}
Here, we can write down the non-zero matrix elements
of $T$:\,\footnote{The two indices of $T$ and one of the
  indices of $U$
  must be that of flavor $\alpha = e, \mu, \tau, s$.
  For simplicity, we denote $T_{ij}~(i,j=1,\cdots,4)$
  and $U_{ij}~(i,j=1,\cdots,4)$
  instead of $T_{\alpha\beta}~(\alpha,\beta = e, \mu, \tau, s)$
  and $U_{\alpha i}~(\alpha = e, \mu, \tau, s;i=1,\cdots,4)$
  in this appendix.}
\begin{align}
\begin{aligned}
  \hspace*{-20mm}
  T_{11} 
&=
\tilde{c}_{14}^2 e^{-i L \varepsilon_{1 -}} 
+ 
\tilde{s}_{14}^2 e^{-i L \varepsilon_{1 +}} \,, 
\\
  \hspace*{-20mm}
T_{14} 
&=
T_{41}
=
\tilde{s}_{14} \tilde{c}_{14} 
\left( e^{-i L \varepsilon_{1 +}} 
- e^{-i L \varepsilon_{1 -}} \right) \,, 
\\
  \hspace*{-20mm}
T_{22} & = T_{33} = 1 \,, 
\\
  \hspace*{-20mm}
T_{44} 
&=
\tilde{s}_{14}^2 e^{-i L \varepsilon_{1 -}} 
+ 
\tilde{c}_{14}^2 e^{-i L \varepsilon_{1 +}} \,,
\end{aligned}
\label{tij}
\end{align}
where $\tilde{s}_{14} = \sin{\tilde\theta_{14}}$, $\tilde{c}_{14} =
\cos{\tilde\theta_{14}}$ and $\varepsilon_{1 \mp}$ is given by
Eq.\,(\ref{epsilon1}).
From Eq.\,(\ref{tij}) we have the following:
\begin{align}
  |T_{11}|^2 &= 1 - \sin^22\tilde\theta_{14}
  \sin^2\left(\frac{\Delta\tilde{E}_1L}{2}\right)
\label{t112}\\
|T_{14}|^2 &= \sin^22\tilde\theta_{14}
  \sin^2\left(\frac{\Delta\tilde{E}_1L}{2}\right)
\tag{\ref{t142}}\\
|T_{41}|^2 &= \sin^22\tilde\theta_{14}
  \sin^2\left(\frac{\Delta\tilde{E}_1L}{2}\right)
\\
|T_{44}|^2 &= 1 - \sin^22\tilde\theta_{14}
  \sin^2\left(\frac{\Delta\tilde{E}_1L}{2}\right)
\end{align}

For later purposes, we keep the dependence of $U$
on $\theta_{14}$ but we use the following
approximation for the three-flavor mixing angles in $U$:
\begin{eqnarray}
  &{\ }&\hspace*{-85mm}
\theta_{13}\simeq 0\,.
\end{eqnarray}
We introduce the product of
matrices $T$ and $U$:
\begin{align}
  V&\equiv TU
  \notag \\
&\simeq\left( \begin{array}{cccc}
T_{11} & T_{12} & T_{13} & T_{14}\\
T_{21} & T_{22} & T_{23} & T_{24} \\
T_{31} & T_{32} & T_{33} & T_{34} \\
T_{41} & T_{42} & T_{43} & T_{44}
\end{array} \right)
\left( \begin{array}{cccc}
c_{14} & 0 & 0 & s_{14}\\
0 & 1 & 0 & 0 \\
0 & 0 & 1 & 0 \\
-s_{14} & 0 & 0 & c_{14}
\end{array} \right)
\notag \\
&\quad\times
\left( \begin{array}{cccc}
c_{12} & s_{12} & 0 & 0\\
-s_{12}c_{23} & c_{12}c_{23} & s_{23} & 0 \\
s_{12}s_{23} & -c_{12}s_{23} & c_{23} & 0 \\
0 & 0 & 0 & 1
\end{array} \right)
\\
&=\left( \begin{array}{cccc}
  c_{12}\,T_{1-}
  & s_{12}\,T_{1-}
  & 0
  &T_{1+} \\
  -s_{12}c_{23}
  & c_{12}c_{23}
  & s_{23}
  & 0 \\
  s_{12}s_{23}
  & -c_{12}s_{23}
  & c_{23}
  & 0 \\
  c_{12}\,T_{4-}
  & s_{12}\,T_{4-}
  & 0
  & T_{4+}
\end{array} \right)
\end{align}
where
\begin{align}
  T_{1-}&\equiv c_{14}\,T_{11}-s_{14}\,T_{14}
  \label{t1-}\\
T_{1+}&\equiv s_{14}\,T_{11}+c_{14}\,T_{14}
  \\
  T_{4-}&\equiv c_{14}\,T_{41}-s_{14}\,T_{44}
  \label{t4-}\\
T_{4+}&\equiv s_{14}\,T_{41}+c_{14}\,T_{44}
\end{align}
Then, from Eq.\,(\ref{P7}), we obtain the
formulae of the oscillation probabilities:
\begin{align}
 &\quad P(\nu_\alpha \to \nu_\beta)
\notag \\
&=
\sum_i
|V_{\beta i}|^2 |U_{\alpha i}|^2 \notag \\
&\simeq\left[
  \left( \begin{array}{cccc}
    c_{12}^2 |T_{1-}|^2 &s_{12}^2 |T_{1-}|^2 &0 &|T_{1+}|^2 \\
    s_{12}^2 c_{23}^2 &c_{12}^2 c_{23}^2 &s_{23}^2 &0 \\
    s_{12}^2 s_{23}^2 &c_{12}^2 s_{23}^2 &c_{23}^2 &0 \\
    c_{12}^2 |T_{4-}|^2 &s_{12}^2 |T_{4-}|^2 &0 & |T_{4+}|^2
\end{array} \right)
\left( \begin{array}{cccc}
c_{12}^2 c_{14}^2 & s_{12}^2 c_{23}^2 & s_{12}^2 s_{23}^2 &c_{12}^2 s_{14}^2\\
s_{12}^2 c_{14}^2 & c_{12}^2 c_{23}^2 & c_{12}^2 s_{23}^2 &s_{12}^2 s_{14}^2\\
0&s_{23}^2&c_{23}^2&0\\
s_{14}^2 & 0 & 0 & c_{14}^2
\end{array} \right) \right]_{\beta\alpha}\,.
\end{align}
Each probability is as follows:
\begin{align}
  \hspace*{-10mm}
P(\nu_e \to \nu_e)
& =
c_{14}^2\left( s_{12}^4 + c_{12}^4 \right)
\,|T_{1-}|^2 + s_{14}^2\,|T_{1+}|^2\,,
\\
  \hspace*{-10mm}
P(\nu_e \to \nu_\mu)
& =
2 c_{14}^2 s_{12}^2 c_{12}^2 c_{23}^2 \,,
\\
  \hspace*{-10mm}
P(\nu_e \to \nu_\tau)
& =
2 c_{14}^2 s_{12}^2 c_{12}^2 s_{23}^2 \,,
\\ 
  \hspace*{-10mm}
P(\nu_e \to \nu_s)
& =
c_{14}^2 \left( s_{12}^4 + c_{12}^4 \right)\,|T_{4-}|^2
+ s_{14}^2 \,|T_{4+}|^2
\ ;
\end{align}
and
\begin{eqnarray}
  &{\ }&\hspace*{-40mm}
P(\nu_\mu \to \nu_e)
 =
2 s_{12}^2 c_{12}^2 c_{23}^2\,|T_{1-}|^2 \,,
\\
  &{\ }&\hspace*{-40mm}
P(\nu_\mu \to \nu_\mu)
 =
\left( s_{12}^4 + c_{12}^4 \right)
c_{23}^4
+
s_{23}^4 \,,
\\
  &{\ }&\hspace*{-40mm}
P(\nu_\mu \to \nu_\tau)
 =
\left( s_{12}^4 + c_{12}^4 + 1\right) s_{23}^2 c_{23}^2
\,,
\\ 
  &{\ }&\hspace*{-40mm}
P(\nu_\mu \to \nu_s)
 =
2 s_{12}^2 c_{12}^2 c_{23}^2\,|T_{4-}|^2 \,.
\end{eqnarray}

It is known that at the production source of astrophysical neutrinos,
such as active galactic nuclei, the ratios of the flux of
$\nu_e$, $\nu_\mu$ and $\nu_\tau$ is
1:2:0.  Moreover, it is expected that there is no
sterile neutrino at the source.  Denoting the flux of astrophysical
neutrinos at the source as $F^{0}(\nu_{\alpha}) \ (\alpha = e, \mu,
\tau, s)$, we have $F^{0}(\nu_e) : F^{0}(\nu_\mu) : F^{0}(\nu_\tau) :
F^{0}(\nu_s) = 1:2:0:0$.  From Eq.\,(\ref{osc-param3}), to a good
approximation, we can assume that $\theta_{23} \simeq \pi / 4$, so
that $s_{23} \simeq c_{23} \simeq 1/\sqrt{2}$.  Thus we can estimate
the flux of $\nu_e$, $\nu_\mu$, $\nu_\tau$ and $\nu_s$,
which is observed at IceCube is given in the unit of $F^{0}(\nu_e)$
by:
\begin{align}
F(\nu_e)
& = 
P(\nu_e \to \nu_e)
+
2 P(\nu_\mu \to \nu_e) \notag \\
& \simeq
\left\{c_{14}^2 \left(
s_{12}^4 
+
c_{12}^4\right)
+
2 s_{12}^2 c_{12}^2
\right\}\,|T_{1-}|^2
+s_{14}^2 |T_{1+}|^2\,,
\label{fluxe}\\
F(\nu_\mu)
& = 
P(\nu_e \to \nu_\mu)
+
2 P(\nu_\mu \to \nu_\mu) \notag \\
& \simeq
c_{14}^2 s_{12}^2 c_{12}^2
+
\frac{1}{2} \left( s_{12}^4 + c_{12}^4 + 1\right)
\,,
\\
F(\nu_\tau)
& = 
P(\nu_e \to \nu_\tau)
+
2 P(\nu_\mu \to \nu_\tau) \notag \\
& \simeq 
c_{14}^2 s_{12}^2 c_{12}^2
+
\frac{1}{2} \left( s_{12}^4 + c_{12}^4 + 1\right)
\,,
\\
F(\nu_s)
& = 
P(\nu_e \to \nu_s)
+
2 P(\nu_\mu \to \nu_s) \notag \\
& \simeq
\left\{ c_{14}^2 \left(
s_{12}^4 
+
c_{12}^4\right)
+
2 s_{12}^2 c_{12}^2
\right\}\,|T_{4-}|^2
+s_{14}^2 |T_{4+}|^2 \,.
\label{fluxs}
\end{align}

If we use approximation $\theta_{14}\simeq 0$
except in $T_{i\pm}~(i=1,4)$
while keeping $\tilde\theta_{14}\ne 0$ in each
flux, from Eqs.\,(\ref{t1-}), (\ref{t4-}),
(\ref{t112}) and (\ref{t142}), we have
\begin{align}
|T_{1-}|^2&\simeq |T_{11}|^2=1 - \sin^22\tilde\theta_{14}
  \sin^2\left(\frac{\Delta\tilde{E}_1L}{2}\right)
  \\
|T_{4-}|^2&\simeq |T_{41}|^2= \sin^22\tilde\theta_{14}
  \sin^2\left(\frac{\Delta\tilde{E}_1L}{2}\right)
\end{align}
Hence, the ratio of each flavor of neutrino can be
approximated as
\begin{align}
F(\nu_e) 
:
F(\nu_\mu) 
:
F(\nu_\tau)
:
F(\nu_s) 
&\simeq 1-|T_{14}|^2 : 1 : 1: |T_{14}|^2\,,
\tag{\ref{ratio1}}
\end{align}
where $|T_{14}|^2$ is defined in Eq.\,(\ref{t142})
and $|T_{14}|^2$ has nontrivial dependence
on the neutrino energy $E$, especially near
the resonance region $E {\color{black}= \Delta{m}^2_{41}\cos{2\theta_{14}}/\{2(A_e - A_s)\}}
\sim \Delta m^2_{41}/\{2(A_e-A_s)\}${\color{black},
where we have assumed $|\theta_{14}|\ll 1$}.

\section{Derivation of Eq.\,(\ref{ratio2}) in the case of
  $\theta_{14} = 0$, $\theta_{24}\ne 0$, and $\theta_{34}\ne 0$}
\label{appendix-d}

In this Appendix, we discuss oscillations of
astrophysical neutrinos for
$\theta_{14} = 0$, $\theta_{24} \ne 0$ and $\theta_{34} \ne 0$.  As in
Subsubsection \ref{Subsubsec3.1.2}, it is useful to discuss in terms
of the new mixing angles in Eqs.\,(\ref{phi12}), (\ref{phi13}),
(\ref{13.1}).  According to Eq.\,(\ref{H2.4}) in Subsubsection
\ref{Subsubsec3.1.2}, we can explicitly express the transition matrix $T$:
\begin{align}
T
=
\left( \begin{array}{cccc}
e^{-i L A_e}        & 0 & 0 & 0 \\
0 & T_{22} & T_{23} & T_{24} \\
0 & T_{32} & T_{33} & T_{34} \\
0 & T_{42} & T_{43} & T_{44}
\end{array} \right) \,. 
\end{align} 
Here, we denote the submatrix of $T$ by $T_{0}$:
\begin{align}
T_{0}
& =
\left( \begin{array}{ccc}
T_{22} & T_{23} & T_{24} \\
T_{32} & T_{33} & T_{34} \\
T_{42} & T_{43} & T_{44}
\end{array} \right) \notag \\
& =
R \,
e^{i \varphi_{23} \lambda_7}
e^{i \tilde\varphi_{13} \lambda_5}
\left( \begin{array}{ccc}
e^{-i L \varepsilon_{2 -}} & 0 & 0 \\   
0                               & 1 & 0 \\   
0 & 0 & e^{-i L \varepsilon_{2 +}}    
\end{array} \right) 
e^{-i \tilde\varphi_{13} \lambda_5}
e^{-i \varphi_{23} \lambda_7}
R \,,
\end{align}
where
\begin{align}
\begin{aligned}
T_{22} 
&= 
\left( \tilde{S}_{13}^2 
e^{-i L \varepsilon_{2 -}}
+
\tilde{C}_{13}^2 
e^{-i L \varepsilon_{2 +}} \right)
C_{23}^2
+ 
S_{23}^2 \,, 
\\
T_{23} 
&=
T_{32}
= 
\left( \tilde{S}_{13}^2 
e^{-i L \varepsilon_{2 -}}
+
\tilde{C}_{13}^2
e^{-i L \varepsilon_{2 +}} 
- 1 \right)
S_{23} C_{23} \,, 
\\
T_{24} 
&=
T_{42}
= 
\left( e^{-i L \varepsilon_{2 +}}
-
e^{-i L \varepsilon_{2 -}} \right)
\tilde{S}_{13} \tilde{C}_{13} C_{23} \,, 
\\
T_{33} 
&= 
\left( \tilde{S}_{13}^2 
e^{-i L \varepsilon_{2 -}}
+
\tilde{C}_{13}^2 
e^{-i L \varepsilon_{2 +}} \right)
S_{23}^2
+ 
C_{23}^2 \,, 
\\
T_{34} 
&=
T_{43}
=
\left( e^{-i L \varepsilon_{2 +}}
-
e^{-i L \varepsilon_{2 -}} \right)
\tilde{S}_{13} \tilde{C}_{13} S_{23} \,, 
\\
T_{44} 
&=
\tilde{C}_{13}^2
e^{-i L \varepsilon_{2 -}}
+
\tilde{S}_{13}^2
e^{-i L \varepsilon_{2 +}} \,.
\end{aligned}
\label{t0}
\end{align}
From Eqs.\,(\ref{t0}),
we have the following:
\begin{align}
|T_{22}|^2
&= S_{23}^4 + C_{23}^4 \left\{1 - \sin^22\tilde\varphi_{13}\,
\sin^2\left(\frac{\Delta\tilde{E}_2 L}{2}\right) \right\}
\nonumber\\
&\quad + 2 C_{23}^2 S_{23}^2 \,\mbox{\rm Re}
\left[e^{-i(\Delta E -A_s)L/2}\,
  \left\{\cos\left(\frac{\Delta\tilde{E}_2 L}{2}\right)
  -i\cos2\tilde\varphi_{13}\sin\left(\frac{\Delta\tilde{E}_2 L}{2}\right)
  \right\}\right] \,,
\\
|T_{23}|^2
&=|T_{32}|^2
\nonumber\\
&=C_{23}^2 S_{23}^2 \,\left\{
2- \sin^22\tilde\varphi_{13}\,
\sin^2\left(\frac{\Delta\tilde{E}_2 L}{2}\right) \right\}
\nonumber\\
&\quad - 2 C_{23}^2 S_{23}^2 \,\mbox{\rm Re}
\left[e^{-i(\Delta E -A_s)L/2}\,
  \left\{\cos\left(\frac{\Delta\tilde{E}_2 L}{2}\right)
  -i\cos2\tilde\varphi_{13}\sin\left(\frac{\Delta\tilde{E}_2 L}{2}\right)
  \right\}\right] \,,
\\
|T_{24}|^2
&=
|T_{42}|^2
\nonumber\\
&= C_{23}^2 \sin^22\tilde\varphi_{13}\,
\sin^2\left(\frac{\Delta\tilde{E}_2 L}{2}\right)\,, 
\\
|T_{33}|^2
&= C_{23}^4 + S_{23}^4 \left\{1 - \sin^22\tilde\varphi_{13}\,
\sin^2\left(\frac{\Delta\tilde{E}_2 L}{2}\right) \right\}
\nonumber\\
&\quad + 2 C_{23}^2 S_{23}^2 \,\mbox{\rm Re}
\left[e^{-i(\Delta E -A_s)L/2}\,
  \left\{\cos\left(\frac{\Delta\tilde{E}_2 L}{2}\right)
  -i\cos2\tilde\varphi_{13}\sin\left(\frac{\Delta\tilde{E}_2 L}{2}\right)
  \right\}\right] \,,
\\
|T_{34}|^2
&=
|T_{43}|^2
\nonumber\\
&= S_{23}^2 \sin^22\tilde\varphi_{13}\,
\sin^2\left(\frac{\Delta\tilde{E}_2 L}{2}\right) \,, 
\\
|T_{44}|^2
&= 1 - \sin^22\tilde\varphi_{13}\,
\sin^2\left(\frac{\Delta\tilde{E}_2 L}{2}\right)\,.
\tag{\ref{t442}}
\end{align}
Here, we have defined
$\tilde{S}_{13} = \sin{\tilde\varphi_{13}}$, 
$\tilde{C}_{13} = \cos{\tilde\varphi_{13}}$,
$S_{23} = \sin{\varphi_{23}}$,
$C_{23} = \cos{\varphi_{23}}$
and
$\varepsilon_{2 \mp}$ is given by Eq.\,(\ref{epsilon2.1}).

Then, according to Eq.\,(\ref{P7}), using approximation $\theta_{13}\simeq 0$,
we can obtain the formulae of
oscillation probabilities:
\begin{align}
P(\nu_e \to \nu_e)
& = 1 - 2 s_{12}^2 c_{12}^2 \,,
\\
P(\nu_e \to \nu_\mu)
& =
2 |T_{22}|^2
\cdot
s_{12}^2 c_{12}^2 c_{23}^2  
-
4 \mbox{\rm Re}\,( T_{22} T_{23}^\ast ) 
\cdot
s_{12}^2 c_{12}^2 s_{23} c_{23}
+
2 |T_{23}|^2
\cdot
s_{12}^2 c_{12}^2 s_{23}^2 \,,
\\
P(\nu_e \to \nu_\tau)
& =
2 |T_{32}|^2
\cdot
s_{12}^2 c_{12}^2 c_{23}^2  
-
4 \mbox{\rm Re}\,( T_{32} T_{33}^\ast ) 
\cdot
s_{12}^2 c_{12}^2 s_{23} c_{23}
+
2 |T_{33}|^2
\cdot
s_{12}^2 c_{12}^2 s_{23}^2 \,,
\\ 
P(\nu_e \to \nu_s)
& =
2 |T_{42}|^2
\cdot
s_{12}^2 c_{12}^2 c_{23}^2  
-
4 \mbox{\rm Re}\,( T_{42} T_{43}^\ast ) 
\cdot
s_{12}^2 c_{12}^2 s_{23} c_{23}
+
2 |T_{43}|^2
\cdot
s_{12}^2 c_{12}^2 s_{23}^2 \ ;
\end{align}
and
\begin{align}
P(\nu_\mu \to \nu_e)
= {} &
2 s_{12}^2 c_{12}^2 c_{23}^2 \,,
\\
P(\nu_\mu \to \nu_\mu)
= {} &
|T_{22}|^2
\cdot
\left[ 
\left( s_{12}^4 + c_{12}^4 \right)
c_{23}^4
+
s_{23}^4 
\right] \notag \\
& -
2 \mbox{\rm Re}\,( T_{22} T_{23}^\ast ) 
\cdot
\left[
\left( s_{12}^4 + c_{12}^4 \right)
s_{23} c_{23}^3
-
s_{23}^3 c_{23}
\right] \notag \\
& +
|T_{23}|^2
\cdot
\left[
\left( s_{12}^4 + c_{12}^4 \right)
s_{23}^2 c_{23}^2
+
s_{23}^2 c_{23}^2
\right] \,,
\\
P(\nu_\mu \to \nu_\tau)
= {} &
|T_{32}|^2
\cdot
\left[ 
\left( s_{12}^4 + c_{12}^4 \right)
c_{23}^4
+
s_{23}^4 
\right] \notag \\
& -
2 \mbox{\rm Re}\,( T_{32} T_{33}^\ast ) 
\cdot
\left[
\left( s_{12}^4 + c_{12}^4 \right)
s_{23} c_{23}^3
-
s_{23}^3 c_{23}
\right] \notag \\
& +
|T_{33}|^2
\cdot
\left[
\left( s_{12}^4 + c_{12}^4 \right)
s_{23}^2 c_{23}^2
+
s_{23}^2 c_{23}^2
\right] \,,
\\ 
P(\nu_\mu \to \nu_s)
= {} &
|T_{42}|^2
\cdot
\left[ 
\left( s_{12}^4 + c_{12}^4 \right)
c_{23}^4
+
s_{23}^4 
\right] \notag \\
& -
2 \mbox{\rm Re}\,( T_{42} T_{43}^\ast ) 
\cdot
\left[
\left( s_{12}^4 + c_{12}^4 \right)
s_{23} c_{23}^3
-
s_{23}^3 c_{23}
\right] \notag \\
& +
|T_{43}|^2
\cdot
\left[
\left( s_{12}^4 + c_{12}^4 \right)
s_{23}^2 c_{23}^2
+
s_{23}^2 c_{23}^2
\right] \,.
\end{align}
As in Subsubsection \ref{Subsubsec3.2.1}, using approximation $\theta_{23}\simeq \pi/4$,
we obtain the flux of
$\nu_e$, $\nu_\mu$ and $\nu_s$ observed at IceCube:
\begin{align}
F(\nu_e)
= {} & 
P(\nu_e \to \nu_e)
+
2 P(\nu_\mu \to \nu_e) \notag \\
\simeq {} &
s_{12}^4 
+
c_{12}^4 
+
4 s_{12}^2 c_{12}^2 c_{23}^2 \notag \\
\simeq {} &
1 \,,
\\
F(\nu_\mu)
= {} & 
P(\nu_e \to \nu_\mu)
+
2 P(\nu_\mu \to \nu_\mu) \notag \\
\simeq {} &
|T_{22}|^2
\cdot
\left[ 
2 s_{12}^2 c_{12}^2 c_{23}^2
+
2 \left( s_{12}^4 + c_{12}^4 \right)
c_{23}^4
+
2 s_{23}^4 
\right] \notag \\
& -
4 \mbox{\rm Re}\,( T_{22} T_{23}^\ast ) 
\cdot
\left[
s_{12}^2 c_{12}^2 s_{23} c_{23}
+
\left( s_{12}^4 + c_{12}^4 \right)
s_{23} c_{23}^3
-
s_{23}^3 c_{23} 
\right] \notag \\
& +
|T_{23}|^2
\cdot
\left[ 
2 s_{12}^2 c_{12}^2 s_{23}^2
+
2 \left( s_{12}^4 + c_{12}^4 \right)
s_{23}^2 c_{23}^2
+
2 s_{23}^2 c_{23}^2 
\right] \notag \\
\simeq {} &
|T_{22}|^2 + |T_{23}|^2 \,,
\\
F(\nu_\tau)
= {} & 
P(\nu_e \to \nu_\tau)
+
2 P(\nu_\mu \to \nu_\tau) \notag \\
\simeq {} &
|T_{32}|^2
\cdot
\left[ 
2 s_{12}^2 c_{12}^2 c_{23}^2
+
2 \left( s_{12}^4 + c_{12}^4 \right)
c_{23}^4
+
2 s_{23}^4 
\right] \notag \\
& -
4 \mbox{\rm Re}\,( T_{32} T_{33}^\ast ) 
\cdot
\left[
s_{12}^2 c_{12}^2 s_{23} c_{23}
+
\left( s_{12}^4 + c_{12}^4 \right)
s_{23} c_{23}^3
-
s_{23}^3 c_{23} 
\right] \notag \\
& +
|T_{33}|^2
\cdot
\left[ 
2 s_{12}^2 c_{12}^2 s_{23}^2
+
2 \left( s_{12}^4 + c_{12}^4 \right)
s_{23}^2 c_{23}^2
+
2 s_{23}^2 c_{23}^2 
\right] \notag \\
\simeq {} &
|T_{32}|^2 + |T_{33}|^2 \,,
\\
F(\nu_s)
= {} & 
P(\nu_e \to \nu_s)
+
2 P(\nu_\mu \to \nu_s) \notag \\
\simeq {} &
|T_{42}|^2
\cdot
\left[ 
2 s_{12}^2 c_{12}^2 c_{23}^2
+
2 \left( s_{12}^4 + c_{12}^4 \right)
c_{23}^4
+
2 s_{23}^4 
\right] \notag \\
& -
4 \mbox{\rm Re}\,( T_{42} T_{43}^\ast ) 
\cdot
\left[
s_{12}^2 c_{12}^2 s_{23} c_{23}
+
\left( s_{12}^4 + c_{12}^4 \right)
s_{23} c_{23}^3
-
s_{23}^3 c_{23} 
\right] \notag \\
& +
|T_{43}|^2
\cdot
\left[ 
2 s_{12}^2 c_{12}^2 s_{23}^2
+
2 \left( s_{12}^4 + c_{12}^4 \right)
s_{23}^2 c_{23}^2
+
2 s_{23}^2 c_{23}^2 
\right] \notag \\
\simeq {} &
|T_{42}|^2 + |T_{43}|^2 \,.
\end{align}
Hence the flavor ratio is given by
\begin{align}
{} &
F(\nu_e) 
: 
F(\nu_\mu) 
: 
F(\nu_\tau)
:
F(\nu_s) 
\notag \\
\simeq {} &
1
:
1 - |T_{24}|^2
:
1 - |T_{34}|^2
:
1 - |T_{44}|^2
\notag \\
= {} &
1
:
1 - C_{23}^2 \,(1 - |T_{44}|^2)
:
1 - S_{23}^2 \,(1 - |T_{44}|^2)
:
1 - |T_{44}|^2
\,,
\tag{\ref{ratio2}}
\end{align}
where $|T_{44}|^2$ is defined in Eq.\,(\ref{t442})
and $|T_{44}|^2$ has nontrivial dependence
on the neutrino energy $E$, especially near
the resonance region $E{\color{black}=-\Delta m^2_{41}\cos2\varphi_{13}/(2A_s)}
\sim \Delta m^2_{41}/(2A_s)${\color{black}, where we have assumed $|\pi/2-\varphi_{13}|\ll 1$}.

\section{The behavior of astrophysical $\nu_e$
as $\theta_{14}\to \pi/4$}
\label{appendix-e}
If $\theta_{14}$ is very close to $\pi/4$ and
$|\Delta E| \ll | A_e - A_s|$,
from Eqs.\,(\ref{theta14tilde}) and (\ref{etilde1}), we obtain
\begin{align}
\cos2\tilde\theta_{14}=
\frac{\Delta E \cos{2\theta_{14}} - A_e + A_s}{\Delta\tilde E_1} \simeq -1\,,
\end{align}
which indicates
\begin{align}
  \tilde\theta_{14}\simeq \frac{\pi}{2}\,.
\end{align}
From Eq.\,(\ref{t1-}), we obtain
\begin{align}
  T_{1-}&\simeq \frac{1}{\sqrt{2}}\,\left(T_{11}-T_{14}\right)
\end{align}
We can now calculate the factor $|T_{1-}|^2$ in Eq.\,(\ref{fluxe})
as $\theta_{14}\to \pi/4$ as follows.  
\begin{align}
  |T_{1-}|^2&\simeq \frac{1}{2}\,\left|T_{11}-T_{14}\right|^2
  \nonumber\\
  &= \frac{1}{2}\,\left\{|T_{11}|^2+|T_{14}|^2
  -2\,\mbox{\rm Re}\,\left(T_{11}T_{14}^\ast\right)\right\}
  \nonumber\\
  &\simeq\frac{1}{2}
\label{t1-limit}
\end{align}
where we used the following properties and
Eqs.\,(\ref{t112}) and (\ref{t142}):
\begin{align}
  T_{11}&= e^{-i(\Delta E+A_e+A_s)L/2}
    \left\{\cos\left(\frac{\Delta\tilde{E}_1L}{2}\right)
    +i\cos2\tilde\theta_{14}\sin\left(\frac{\Delta\tilde{E}_1L}{2}\right)
    \right\}
    \\
    T_{14}&= -ie^{-i(\Delta E+A_e+A_s)L/2}\sin2\tilde\theta_{14}
    \sin\left(\frac{\Delta\tilde{E}_1L}{2}\right)\simeq 0 \,.
\end{align}
Eq.\,(\ref{t1-limit}) implies that
the oscillation probability of $\nu_e$ flux
is independent of the neutrino energy $E$ and
baseline length $L$ in the Earth.

\end{document}